\documentclass[aps,prresearch,twocolumn,nofootinbib, superscriptaddress]{revtex4-2}

\usepackage{amsmath,amssymb,amsthm,bm}
\usepackage{physics}
\usepackage{graphicx}
\usepackage{xcolor}

\usepackage{hyperref}
\usepackage{mathtools}
\theoremstyle{plain}

\usepackage{tikz}
\usetikzlibrary{arrows.meta, positioning, fit}

\begin{document}
\title{From Analytic Structure to Quantum Complexity:
Walsh--Pauli Representations of Continuum Operators}

\author{Raphael De Sousa}
\affiliation{Rudolf Peierls Centre for Theoretical Physics,
University of Oxford, Oxford OX1 3PU, United Kingdom}
\affiliation{National Institute for Theoretical and Computational Sciences (NITheCS), South Africa}

\author{Partha Nandi}
\affiliation{National Institute for Theoretical and Computational Sciences (NITheCS), South Africa}

\author{Francesco Petruccione}
\affiliation{National Institute for Theoretical and Computational Sciences (NITheCS), South Africa}
\affiliation{School of Data Science and Computational Thinking,
Stellenbosch University, Stellenbosch 7600, South Africa}
\affiliation{School of Physics,
Stellenbosch University, Stellenbosch 7600, South Africa}

\begin{abstract}

How does continuum operator structure appear in a finite qubit
register? We address this question for analytic functions of diagonal
operators represented in a binary basis, focusing on the
Jordan--Lee--Preskill (JLP) momentum operator, by developing an
operator-level generating-function framework for their Walsh--Pauli
spectra. The construction determines the spectrum analytically and
reveals exact support and parity selection rules, together with a
nontrivial intra-shell hierarchy induced by the binary encoding.
This provides a systematic characterization of Walsh
compressibility and its relation to Pauli locality. As a controlled
application, we consider generalized uncertainty-principle (GUP)
kinematics as a tunable nonlinear odd deformation of momentum, showing
how higher-order momentum terms redistribute spectral weight into
progressively higher Pauli-weight sectors.

\end{abstract}
\maketitle

\section{INTRODUCTION}

Quantum simulation provides a natural setting in which continuum quantum
dynamics can be represented on finite-dimensional quantum processors
\cite{Feynman1982,Lloyd1996}. A central challenge in digital quantum
simulation is therefore to understand how the structure of a continuum
operator is translated into a finite qubit register. Although a generic
operator acting on an $n_q$-qubit Hilbert space may require an exponentially
large number of Pauli strings, physically structured operators can occupy
a much smaller and organized subset of the operator space. When the
corresponding Pauli spectrum is concentrated in low-weight sectors, the
resulting dynamics can be represented using relatively low-weight Pauli
strings, which can be advantageous for quantum-circuit implementation.
This motivates a more specific question: \emph{how does the analytical
structure of a continuum momentum operator determine the structure and
Pauli-weight organization of its finite-qubit representation?}

For diagonal operators, this question can be formulated naturally in
terms of Walsh functions. On a binary register, Walsh functions are
closely related to tensor products of Pauli-$Z$ operators, establishing
an exact correspondence between Walsh expansions of functions on a
binary lattice and Pauli-$Z$ expansions of diagonal quantum operators
\cite{Walsh1923,Welch2014,Hadfield:2018yak}. Thus, after discretization
and binary encoding, a diagonal operator can be represented as a
function on the Boolean hypercube. The Walsh--Pauli correspondence itself
is well established. The question of interest here is instead how the
Walsh spectrum of an analytically specified operator can be determined
and organized directly from the analytical form of the operator and the
underlying binary encoding, without first evaluating the function
independently on the full discrete lattice.

Walsh representations have been used extensively in quantum-circuit
synthesis, Hamiltonian simulation, and quantum-state preparation, where
sparse or truncated expansions can reduce the resources required for
quantum implementation
\cite{Welch2014,kane2022efficientquantumimplementation21,
ZylbermanDebbasch2024}. In these approaches, the Walsh coefficients of a
target function are typically obtained from its values on the discrete
lattice and subsequently converted into a circuit representation. Here,
we address the preceding analytical problem: given an analytically
specified function $F(\hat p)$ of a discretized momentum operator, can
its Walsh--Pauli spectrum be obtained directly from the analytical form
of $F$ and the binary structure of the momentum encoding?

We address this problem using the Jordan--Lee--Preskill (JLP)
discretization of a continuum momentum degree of freedom
\cite{JordanLeePreskill2012,JordanLeePreskill2014,Klco_2019,
sinha2025lecturesquantumfieldtheory}. The JLP construction provides a
finite-dimensional representation of continuum degrees of freedom,
with the resulting degrees of freedom encoded in binary qubit
registers; see Ref.~\cite{sinha2025lecturesquantumfieldtheory} for a
recent pedagogical treatment in the context of quantum-field-theory
simulation. The JLP momentum operator is particularly convenient for
our purposes because, in the momentum basis, it is a linear combination
of mutually commuting single-qubit Pauli-$Z$ operators. This structure
allows powers and analytic functions of the momentum operator to be
treated directly in the Walsh--Pauli basis. Although the JLP momentum
operator is our principal realization, the underlying construction
applies more generally to analytic functions of diagonal operators
represented in a binary basis.
The purpose of this work is to develop an operator-level analytical
framework for determining the Walsh--Pauli coefficients of powers of the
JLP momentum operator and, consequently, of general analytic functions
$F(\hat p)$. Rather than evaluating $F$ at all $2^{n_q}$ lattice points
and subsequently performing a Walsh transform, the generating function
constructs the coefficients directly from the operator. This makes the
structure of the resulting spectrum explicit: polynomial degree bounds
the maximum Pauli weight, parity determines which weight sectors are
allowed, and the binary coefficients of the JLP momentum operator
generate a nonuniform hierarchy among Pauli strings within a fixed
weight sector.

For a general analytic function
$F(p)=\sum_m a_m p^m$, the resulting coefficient of a Pauli string
$Z_S$ takes the form
\[
\alpha_S[F]=\sum_m a_m c_{m,S},
\]
where the coefficients $c_{m,S}$ are determined by the JLP momentum
encoding. This separates two sources of structure in the digital
spectrum: the binary encoding determines the organization of the
momentum-power coefficients, while the analytical form of $F$ determines
how different momentum orders combine. The resulting distribution of
spectral weight across Pauli-weight sectors provides a natural basis for
quantifying Walsh compressibility.

The same generating-function structure also leads to a recursive
construction of the relevant coefficients. Because the generating
function factorizes over the qubit register, the coefficients can be
built by incorporating the qubit factors successively and, when
appropriate, restricting the calculation to the Pauli-weight sectors
of interest. This provides a direct analytical route to selected
low-weight coefficients without explicitly constructing the complete
momentum-space data set. The resulting Walsh spectrum can then be used
to characterize the Pauli-weight structure and the resources required
for standard Pauli-rotation synthesis of the corresponding diagonal
evolution.

The relevance of this framework becomes particularly clear for
deformations of canonical quantum kinematics in which the momentum
dependence is replaced by a nonlinear analytic function. Generalized
uncertainty-principle (GUP) models \cite{Maggiore1993}
provide a concrete example of this situation, since the modified
momentum--wave-number relation can be written as a nonlinear function
of the canonical momentum \cite{PhysRevD.109.024028}. Such modified
quantum kinematics can also be viewed geometrically in terms of a
nontrivial momentum-space geometry, with a correspondence between GUP
deformations and quantum mechanics on curved momentum space
\cite{PhysRevD.104.126010, Nandi:2025mco}. Related momentum-dependent modifications of the
canonical commutation relations arise from the Berry--Keating program,
in particular through the Bender--Brody--Müller construction
\cite{PhysRevD.99.026012}. In the quadratic GUP considered here,
this function is odd and analytic in momentum and admits an expansion
in successively higher odd powers. This makes GUP a useful controlled
setting in which to examine how nonlinear momentum structure is
reflected in the Walsh--Pauli spectrum.

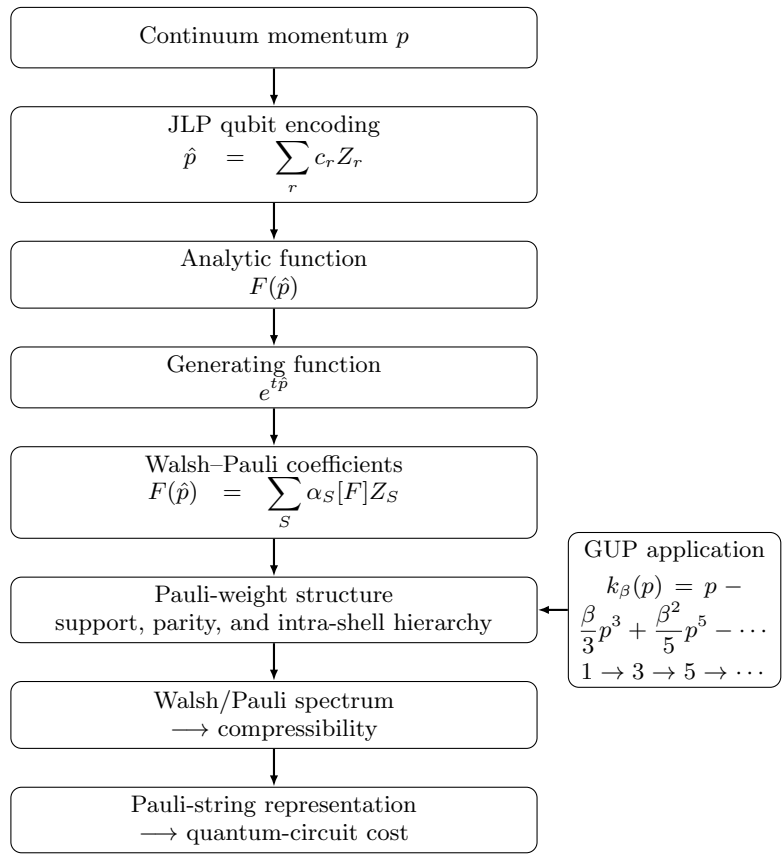
\begin{figure}[t]
\centering
\begin{tikzpicture}[
    node distance=5mm,
    box/.style={
        draw,
        rounded corners,
        align=center,
        text width=0.78\columnwidth,
        minimum height=8mm,
        font=\small
    },
    arrow/.style={
        -{Latex[length=1.5mm]},
        thick
    }
]

\node[box] (p)
{Continuum momentum $p$};

\node[box, below=of p] (jlp)
{JLP qubit encoding\\
$\displaystyle \hat p=\sum_r c_r Z_r$};

\node[box, below=of jlp] (F)
{Analytic function\\
$F(\hat p)$};

\node[box, below=of F] (gen)
{Generating function\\
$\displaystyle e^{t\hat p}$};

\node[box, below=of gen] (coeff)
{Walsh--Pauli coefficients\\
$\displaystyle F(\hat p)=\sum_S\alpha_S[F]Z_S$};

\node[box, below=of coeff] (structure)
{Pauli-weight structure\\
support, parity, and intra-shell hierarchy};

\node[box, below=of structure] (comp)
{Walsh/Pauli spectrum\\
$\longrightarrow$ compressibility};

\node[box, below=of comp] (circuit)
{Pauli-string representation\\
$\longrightarrow$ quantum-circuit cost};

\draw[arrow] (p) -- (jlp);
\draw[arrow] (jlp) -- (F);
\draw[arrow] (F) -- (gen);
\draw[arrow] (gen) -- (coeff);
\draw[arrow] (coeff) -- (structure);
\draw[arrow] (structure) -- (comp);
\draw[arrow] (comp) -- (circuit);

\node[box, right=4mm of structure, text width=0.30\columnwidth]
(gup)
{GUP application\\[1mm]
$\displaystyle
k_\beta(p)
=p-\frac{\beta}{3}p^3
+\frac{\beta^2}{5}p^5-\cdots$
\\[1mm]
$1\rightarrow3\rightarrow5\rightarrow\cdots$};

\draw[arrow] (gup.west) -- (structure.east);

\end{tikzpicture}

\caption{Schematic overview of the continuum-to-qubit construction developed in this work. 
The JLP encoding maps the continuum momentum to a binary qubit operator. 
An analytic function of the momentum operator is decomposed through a generating-function construction into Walsh--Pauli coefficients. 
The resulting spectrum determines the Pauli-weight structure and its compressibility, which are directly relevant for Pauli-string implementation and quantum-circuit resources. 
The quadratic GUP provides a concrete application of the general framework.}
\label{fig:workflow}
\end{figure}

We use GUP here not as a claim of Planck-scale experimental detectability,
but as a tunable family of nonlinear analytic deformations with a
well-defined undeformed limit. Its power-series structure provides a
direct test of the general framework: the successive odd powers of
momentum generate progressively higher Pauli-weight sectors, while
their coefficients determine the amount of spectral weight transferred
among those sectors. The GUP example therefore illustrates how the
analytic structure of a deformation is translated into the Pauli-weight
organization of its finite-qubit representation.

The GUP analysis also illustrates the distinction between structural
and model-dependent features of the digital spectrum. The support and
parity constraints follow from the structure of the momentum powers,
whereas the numerical values of the Walsh coefficients depend on the
Taylor coefficients of the particular deformation. Contributions from
different momentum orders can combine coherently within individual
Pauli strings, so the detailed spectral distribution contains
information beyond the overall strength of the deformation.

The present framework is complementary to Walsh-based
quantum-circuit-synthesis methods \cite{Welch2014}. Those methods address
the conversion of a known Walsh spectrum into a circuit; here we focus
on determining the spectrum itself from the analytical form of the
operator. The resulting organization of the spectrum provides a direct
connection between the JLP encoding, Pauli-weight structure, Walsh
compressibility, and the resources required for the corresponding
digital representation. A schematic overview of the continuum-to-qubit construction and its
connection to Walsh spectra, Pauli-weight structure, compressibility,
and quantum-circuit resources is shown in Fig.~\ref{fig:workflow}.

The scope of this work is restricted to diagonal functions of a single
JLP momentum register. In particular, we do not construct here a
complete finite-dimensional realization of the position--momentum
structure associated with a modified momentum relation, nor do we
address multi-mode or general off-diagonal continuum operators. These
extensions constitute natural directions for future work.

The remainder of the paper is organized as follows. Section~II
introduces the JLP momentum representation and the Walsh--Pauli
formulation. Section~III develops the generating-function formalism and
derives the structural properties of the Walsh coefficients.
Section~IV develops the recursive construction of the Walsh spectrum.
Section~V introduces the shell decomposition and quantifies Walsh
compressibility through high-weight spectral bounds. Section~VI applies
the framework to GUP-deformed spectral functions. Section~VII examines
the universality of the Walsh structure, its dependence on the binary
encoding, and the model dependence of the spectral distribution.
Section~VIII translates the Walsh-spectrum structure into
quantum-circuit resources, including Pauli-rotation counts, locality,
truncation, and elementary-gate costs. Section~IX discusses the main
results, limitations, and possible extensions.

\section{JLP Momentum Operator and Walsh--Pauli Representation}
\label{sec:JLP_Walsh}

We begin by recalling the Jordan--Lee--Preskill (JLP) construction
\cite{JordanLeePreskill2012,JordanLeePreskill2014}, which provides a finite-dimensional
quantum representation of a continuous spatial and momentum degree of
freedom. The continuum is replaced by a finite lattice whose basis
states are encoded in an $n_q$-qubit computational register. The
feature of the JLP construction that is essential for the present work
is that the resulting momentum operator can be expressed as a linear
combination of mutually commuting single-qubit Pauli-$Z$ operators.
This structure provides the algebraic basis for the Walsh analysis
developed below.

\subsection{JLP discretization}

We restrict the spatial coordinate to a finite interval and discretize
it using
\begin{equation}
N=2^{n_q}
\end{equation}
lattice sites,
\begin{equation}
x_\ell=-x_{\max}+\ell\,\delta x,
\qquad
\delta x=\frac{2x_{\max}}{N-1},
\qquad
\ell=0,\ldots,N-1.
\label{eq:x_lattice}
\end{equation}
The lattice index is encoded by the binary expansion
\begin{equation}
\ell
=
\sum_{r=0}^{n_q-1}2^r b_r,
\qquad
b_r\in\{0,1\},
\label{eq:binary_index}
\end{equation}
with
\begin{equation}
b_r=\frac{I-Z_r}{2}.
\label{eq:bit_Z}
\end{equation}
The discrete Fourier transformation then relates the spatial and
momentum representations. For the symmetric JLP construction, the
momentum operator can be written as
\begin{equation}
\hat p
=
\sum_{r=0}^{n_q-1}c_r Z_r ,
\label{eq:p_JLP}
\end{equation}
where
\begin{equation}
c_r
=
-\frac{\pi}{N\delta x}\,2^r .
\label{eq:c_r}
\end{equation}
The overall sign depends on the convention used to order the momentum
basis states and is irrelevant for the structural results below. The
important features are the binary scaling $|c_r|\propto 2^r$ and the
commuting Pauli algebra
\begin{equation}
[Z_r,Z_s]=0,
\qquad
Z_r^2=I .
\label{eq:Z_algebra}
\end{equation}

Equation~\eqref{eq:p_JLP} is the key structural property of the JLP
representation used here. It converts the discretized momentum
operator into a sum of commuting binary degrees of freedom, allowing
powers and analytic functions of $\hat p$ to be organized directly in
the Pauli-$Z$ basis.

More generally, although we focus throughout on the JLP momentum operator, the same construction applies equally to the position operator. Both arise naturally from the binary encoding once the correspondence between Walsh characters and Pauli-$Z$ strings is established. This correspondence extends immediately to constructing operators for arbitrary functions defined on a discretized space, and motivates the Walsh--Pauli formulation introduced in the next section.

\subsection{Walsh--Pauli representation}

The binary encoding identifies the discretized momentum lattice with
the Boolean hypercube
\begin{equation}
\mathbb{B}_{n_q}=\{0,1\}^{n_q}.
\end{equation}
The natural basis functions on this space are the Walsh characters \cite{fine1949, Beauchamp1984},
labelled by subsets
$S\subseteq\{0,\ldots,n_q-1\}$,
\begin{equation}
\chi_S(\ell)
=
(-1)^{\sum_{r\in S}b_r}.
\label{eq:walsh_character}
\end{equation}
Since
\begin{equation}
Z_r|b_r\rangle=(-1)^{b_r}|b_r\rangle,
\end{equation}
each Walsh character is the diagonal matrix element of the Pauli string
\begin{equation}
Z_S
\equiv
\prod_{r\in S}Z_r,
\qquad
Z_{\varnothing}=I .
\label{eq:ZS}
\end{equation}
Thus, for diagonal operators, the Walsh and Pauli-$Z$ bases are
equivalent descriptions of the same binary structure.

The Walsh characters satisfy
\begin{equation}
\frac{1}{N}
\sum_{\ell=0}^{N-1}
\chi_S(\ell)\chi_T(\ell)
=
\delta_{S,T},
\label{eq:walsh_orthogonality}
\end{equation}
so that a function evaluated on the momentum lattice has the exact
Walsh expansion
\begin{equation}
F(p_\ell)
=
\sum_S
\alpha_S[F]\,\chi_S(\ell),
\end{equation}
with
\begin{equation}
\alpha_S[F]
=
\frac{1}{N}
\sum_{\ell=0}^{N-1}
F(p_\ell)\chi_S(\ell).
\label{eq:walsh_coefficients}
\end{equation}
At the operator level,
\begin{equation}
F(\hat p)
=
\sum_S
\alpha_S[F]\,Z_S .
\label{eq:F_p_pauli}
\end{equation}
The Walsh coefficients are therefore directly the coefficients of the
Pauli-$Z$ strings in the diagonal quantum operator\cite{PhysRevA.69.062321}.

The coefficients in Eq.~\eqref{eq:walsh_coefficients} could be obtained
by evaluating $F$ on all $N=2^{n_q}$ momentum points and performing a
discrete Walsh transform. The JLP structure, however, allows the
spectrum to be accessed without treating these exponentially many
values as independent data. Since the operators in
Eq.~\eqref{eq:p_JLP} commute and satisfy $Z_r^2=I$,
\begin{equation}
e^{t\hat p}
=
\prod_{r=0}^{n_q-1}
e^{tc_rZ_r}
=
\prod_{r=0}^{n_q-1}
\left[
\cosh(c_rt)I+
\sinh(c_rt)Z_r
\right].
\label{eq:exp_p}
\end{equation}
Expanding the product in the Pauli basis gives
\begin{equation}
e^{t\hat p}
=
\sum_S g_S(t)Z_S,
\label{eq:generating_expansion}
\end{equation}
where
\begin{equation}
g_S(t)
=
\prod_{r\in S}
\sinh(c_rt)
\prod_{r\notin S}
\cosh(c_rt).
\label{eq:gS}
\end{equation}

Equation~\eqref{eq:gS} provides the generating function from which the
Walsh coefficients of powers of $\hat p$ and, subsequently, of general
analytic functions $F(\hat p)$ can be obtained. The resulting
coefficient-extraction formula and the structural properties of the
Walsh spectrum are developed in the next section.

\section{Exact Walsh Spectrum from the Generating Function}
\label{sec:exact_spectrum}

Using the generating function in Eq.~\eqref{eq:gS}, the Walsh
coefficients of powers of the JLP momentum operator can be obtained
directly by coefficient extraction. Writing
\begin{equation}
\hat p^a
=
\sum_S c_{a,S}Z_S,
\label{eq:pa_power_expansion}
\end{equation}
the Taylor expansion of the generating function gives
\begin{equation}
g_S(t)
=
\sum_{a=0}^{\infty}
\frac{c_{a,S}}{a!}t^a .
\end{equation}
Hence,
\begin{equation}
c_{a,S}
=
\left.
\frac{d^a}{dt^a}g_S(t)
\right|_{t=0}
=
a! \ \text{coeff} \Big(g_S(t) , \ t^a \Big).
\label{eq:exact_coefficient}
\end{equation}
Thus, the Walsh coefficient of every power $\hat p^a$ is determined
analytically from the generating function, without evaluating the
operator on the full $2^{n_q}$-point momentum lattice and performing a
discrete Walsh transform.

The small-$t$ behavior of the same generating function immediately
determines the support of the spectrum. For a subset $S$ with
$w=|S|$,
\begin{equation}
\sinh(c_rt)=c_rt+\mathcal{O}(t^3),
\qquad
\cosh(c_rt)=1+\mathcal{O}(t^2),
\end{equation}
and therefore
\begin{equation}
g_S(t)
=
\left(\prod_{r\in S}c_r\right)t^w
+\mathcal{O}(t^{w+2}).
\label{eq:gS_small_t}
\end{equation}
The first contribution from a Pauli string of weight $w$ can thus
occur only at order $t^w$. Equation~\eqref{eq:exact_coefficient}
implies
\begin{equation}
c_{a,S}=0
\qquad
\text{for}
\qquad
|S|>a .
\label{eq:support}
\end{equation}
Therefore, the maximum Pauli weight appearing in $\hat p^a$ is
\begin{equation}
w_{\max}(\hat p^a)
=
\min(a,n_q).
\label{eq:max_weight}
\end{equation}
The polynomial degree of the momentum function consequently provides
a direct upper bound on the Pauli locality of its Walsh--Pauli
representation.

This support theorem is a standard result in classical Walsh/Fourier analysis: a polynomial of degree $m$ in $\pm1$-valued variables has Walsh support only on subsets of size at most m~\cite{odonnellanalysisbooleanfunctions} and its monomial form has also been derived by integration by parts~\cite{UpperBounds}. In the present quantum-computing setting, the same structure follows directly from the generating function introduced above, which additionally provides the coefficients themselves.

A second selection rule follows from the parity of the generating
function. Since $\sinh$ is odd while $\cosh$ is even,
\begin{equation}
g_S(-t)
=
(-1)^{|S|}g_S(t).
\label{eq:gS_parity}
\end{equation}
The Taylor series of $g_S(t)$ therefore contains only powers of $t$
having the same parity as $|S|$. It follows that
\begin{equation}
c_{a,S}=0
\qquad
\text{unless}
\qquad
a\equiv |S|\pmod{2}.
\label{eq:parity}
\end{equation}
Together with Eq.~\eqref{eq:support}, this gives the complete set of
allowed Pauli weights for a monomial:
\begin{equation}
|S|=a,a-2,a-4,\ldots,
\qquad
0\leq |S|\leq n_q.
\label{eq:allowed_weights}
\end{equation}

The first few cases are summarized in Table~\ref{tab:weight_examples}.
The table illustrates the simultaneous action of the support and
parity constraints without requiring separate calculations for each
power.

This parity theorem is also standard in classical Walsh/Fourier analysis: odd and even functions have support only on odd- and even-weight Walsh functions, respectively~\cite{odonnellanalysisbooleanfunctions}. In the present quantum-computing setting, this parity structure is immediate from the generating function and the linear Pauli-Z representation of $\hat{p}$.

\begin{table}[t]
\centering
\caption{Allowed Pauli weights in the Walsh--Pauli expansion of the
first few powers of the JLP momentum operator (for $n_q\geq5$).}
\label{tab:weight_examples}
\begin{tabular}{c|c}
\hline
Operator & Allowed Pauli weights $w$ \\
\hline
$\hat p$     & $1$ \\
$\hat p^2$   & $0,2$ \\
$\hat p^3$   & $1,3$ \\
$\hat p^4$   & $0,2,4$ \\
$\hat p^5$   & $1,3,5$ \\
\hline
\end{tabular}
\end{table}

The selection rules determine which shells can be populated, but not
how the coefficients are distributed within an allowed shell. That
information follows from the leading term in Eq.~\eqref{eq:gS_small_t}.
For $w=|S|$, the first nonvanishing contribution of Eq.~\eqref{eq:gS_small_t} gives
\begin{equation}
c_{w,S}
=
w!\prod_{r\in S}c_r .
\label{eq:leading_coefficient}
\end{equation}
For the JLP coefficients, $|c_r|\propto2^r$, and hence
\begin{equation}
|c_{w,S}|
\propto
w!\,2^{\sum_{r\in S}r}.
\label{eq:leading_scaling}
\end{equation}
Thus, the Walsh coefficients are generally nonuniform even within a
fixed Pauli-weight shell. Strings containing the more significant
binary digits acquire parametrically larger leading coefficients. This
intra-shell hierarchy will be important when the total spectral weight
of each shell is estimated. A structurally similar result appears in classical Walsh analysis \cite{UpperBounds}.

The monomial result extends directly to an arbitrary analytic spectral
function,
\begin{equation}
F(p)
=
\sum_{a=0}^{\infty}f_a p^a.
\label{eq:F_taylor}
\end{equation}
Using Eq.~\eqref{eq:pa_power_expansion}, its operator representation
takes the form
\begin{equation}
F(\hat p)
=
\sum_S\alpha_S[F]Z_S,
\end{equation}
where
\begin{equation}
\alpha_S[F]
=
\sum_{a=0}^{\infty}
f_a c_{a,S}.
\label{eq:alpha_series}
\end{equation}
The coefficient-extraction relation
Eq.~\eqref{eq:exact_coefficient} then gives the compact expression
\begin{equation}
\alpha_S[F]
=
\left.
F\!\left(\frac{d}{dt}\right)
g_S(t)
\right|_{t=0}.
\label{eq:analytic_generating}
\end{equation}
This establishes the generating-function representation of the Walsh
spectrum for arbitrary analytic functions of the JLP momentum
operator.

The selection rules also extend to general analytic functions. For an
odd function,
\begin{equation}
F(-p)=-F(p),
\end{equation}
only odd powers occur in its Taylor expansion, and therefore
\begin{equation}
\alpha_S[F]=0
\qquad
\text{for even }|S|.
\label{eq:odd_selection}
\end{equation}
Similarly, an even function has support only on even Pauli
weights. More generally, the Taylor structure of $F$ determines which
of the allowed Pauli-weight sectors are populated and with what
relative strength.

For a fixed register, $F(\hat p)$ is determined entirely by its values
on the finite momentum spectrum $\{p_\ell\}$. The role of analyticity is
therefore not to add information at fixed $n_q$, but to provide a common
functional description across register sizes. The same analytic function
generates the Walsh--Pauli spectrum at each $n_q$, allowing its support,
parity, and coefficient hierarchy to be studied systematically as the
resolution is increased.

The generating-function construction thus determines both the individual
Walsh coefficients and the structural organization of the spectrum. We
now turn to its practical evaluation and derive a recursion that builds
the required coefficients directly from the product structure of the
generating function.

\section{Recursive Construction of the Walsh Spectrum}
\label{sec:recursion}
The coefficient-extraction formula in Eq.~\eqref{eq:exact_coefficient}
determines each Walsh coefficient analytically. For a fixed $n_q$-qubit register, with the corresponding JLP
coefficients $\{c_0,\ldots,c_{n_q-1}\}$ specified, the product structure
of the generating function allows these coefficients to be constructed
recursively by incorporating the qubit factors one at a time. Here
$n\leq n_q$ denotes the number of factors already included in the
recursion and should not be interpreted as changing the physical JLP
register, since the coefficients $c_r$ generally change when $n_q$
itself is varied. The recursion adds the corresponding
$\cosh(c_rt)$ or $\sinh(c_rt)$ factor one qubit at a time and thereby
builds the coefficients associated with that chosen set of $c_r$.

To make this structure explicit, let
\begin{equation}
g_S^{(n)}(t)
=
\prod_{r\in S}\sinh(c_rt)
\prod_{\substack{0\leq r<n\\ r\notin S}}
\cosh(c_rt)
\end{equation}
denote the generating function associated with a subset
$S\subseteq\{0,\ldots,n-1\}$ after the first $n$ qubit factors have
been included in the recursion. At the initial step,
\begin{equation}
g_{\varnothing}^{(0)}(t)=1.
\end{equation}
On incorporating the factor associated with qubit $n$, every subset
of $\{0,\ldots,n\}$ either excludes or includes the new index. The corresponding generating functions obey
\begin{align}
g_S^{(n+1)}(t)
&=
g_S^{(n)}(t)\cosh(c_nt),
&& n\notin S,
\label{eq:recursion_absent}
\\
g_{S\cup\{n\}}^{(n+1)}(t)
&=
g_S^{(n)}(t)\sinh(c_nt).
\label{eq:recursion_present}
\end{align}
Thus, the recursion has two branches: the $\cosh$ branch preserves the
existing subset, while the $\sinh$ branch adds the new index.

For the intermediate $n$-factor construction, write
\begin{equation}
\left(
\sum_{r=0}^{n-1}c_r Z_r
\right)^a
=
\sum_{S\subseteq\{0,\ldots,n-1\}}
c_{a,S}^{(n)}Z_S.
\end{equation}
coefficient extraction gives
\begin{equation}
c_{a,S}^{(n)}
=
\left.
\frac{d^a}{dt^a}
g_S^{(n)}(t)
\right|_{t=0}.
\label{eq:recursive_coeff}
\end{equation}
The recursion can therefore be implemented directly at the level of
the generating functions and truncated to the Taylor order required
for a given polynomial power.

For practical evaluation, it is convenient to work with the Taylor
coefficients rather than with the full functions. Write
\begin{equation}
g_S^{(n)}(t)
=
\sum_{m=0}^{a}
\frac{c_{m,S}^{(n)}}{m!}t^m
+\mathcal{O}(t^{a+1}),
\end{equation}
and use
\begin{align}
\cosh(c_nt)
&=
\sum_{j=0}^{\lfloor a/2\rfloor}
\frac{c_n^{2j}}{(2j)!}t^{2j}
+\mathcal{O}(t^{a+2}),
\\
\sinh(c_nt)
&=
\sum_{j=0}^{\lfloor(a-1)/2\rfloor}
\frac{c_n^{2j+1}}{(2j+1)!}t^{2j+1}
+\mathcal{O}(t^{a+2}).
\end{align}
Multiplication with the two branches in
Eqs.~\eqref{eq:recursion_absent} and
\eqref{eq:recursion_present} then gives the coefficient recursions
\begin{align}
c_{a,S}^{(n+1)}
&=
\sum_{j=0}^{\lfloor a/2\rfloor}
\binom{a}{2j}
c_n^{2j}
c_{a-2j,S}^{(n)},
\label{eq:coefficient_recursion_absent}
\\
c_{a,S\cup\{n\}}^{(n+1)}
&=
\sum_{j=0}^{\lfloor(a-1)/2\rfloor}
\binom{a}{2j+1}
c_n^{2j+1}
c_{a-2j-1,S}^{(n)}.
\label{eq:coefficient_recursion_present}
\end{align}
These relations generate the coefficients of $\hat p^a$ from lower
recursion depths and lower Taylor orders. They also preserve the
selection rules derived in Sec.~\ref{sec:exact_spectrum}: the
$\cosh$ branch changes the Taylor order by an even integer, whereas
the $\sinh$ branch changes it by an odd integer. The support and parity
structure therefore emerges automatically from the recursion rather
than being imposed separately.

The same construction can also be organized by Pauli weight. Grouping
the generating functions according to $w=|S|$, we define the set of generating
functions associated with weight-$w$ subsets after $n$ qubits by
\begin{equation}
\mathcal{G}_{w}^{(n)}
=
\left\{
g_S^{(n)}(t):\,S\subseteq\{0,\ldots,n-1\},
\ |S|=w
\right\}.
\end{equation}
Under the recursion, the $\cosh$ branch preserves the Pauli weight, whereas the $\sinh$ branch maps $w\to w+1$.
For a fixed polynomial order $a$, the support result of
Sec.~\ref{sec:exact_spectrum} restricts the relevant sectors to
\begin{equation}
w\leq a,
\end{equation}
so that only a finite set of Pauli-weight sectors needs to be retained
when the polynomial degree is fixed.

This restriction is particularly useful when only a limited part of the Walsh spectrum is required. The recursion can then be restricted to the relevant Taylor orders and Pauli-weight sectors, avoiding construction of the complete momentum lattice and the full Walsh spectrum. This provides a direct way to target the coefficients needed for a chosen approximation.

Using the coefficients generated by Eqs.~\eqref{eq:coefficient_recursion_absent} and \eqref{eq:coefficient_recursion_present}, together with Eq.~\eqref{eq:alpha_series}, we extend the results from monomials to analytic functions via Eq.~\eqref{eq:F_taylor}. In practice, the series is truncated to the order required by the desired accuracy. The recursion thus separates the universal binary structure encoded in the coefficients $c_r$ from the particular
analytic function specified by the coefficients $f_a$.

For small registers, the recursively generated coefficients can be
checked directly against the discrete Walsh coefficients obtained from
Eq.~\eqref{eq:walsh_coefficients}. Agreement to numerical precision
provides a direct validation of the recursive construction.

The recursion therefore provides the computational realization of the
generating-function framework: the latter determines the exact
structure of the Walsh spectrum, while the former provides a
systematic way of evaluating the required coefficients. In the following section, we use these
coefficients to determine the shell-resolved spectral weight and
derive quantitative bounds on the high-weight Walsh tail.

\section{Walsh Compressibility and High-Weight Bounds}
\label{sec:compressibility}

The exact Walsh coefficients obtained in
Sec.~\ref{sec:exact_spectrum}, together with their recursive
construction in Sec.~\ref{sec:recursion}, allow us to quantify how the
spectral weight is distributed over Pauli-weight sectors. We focus here
on the high-weight tail of the Walsh spectrum and derive bounds that
relate its suppression to the analytic structure of the underlying
analytic function.

For functions of the momentum operator in the Pauli basis Eq.~\eqref{eq:F_p_pauli} we define the weight-$w$ spectral content by
\begin{equation}
W_w[F]
=
\sum_{|S|=w}
|\alpha_S[F]|^2 .
\label{eq:shell_weight}
\end{equation}
The total Walsh norm is
\begin{equation}
\mathcal{N}_F
=
\sum_{w=0}^{n_q}W_w[F]
=
\sum_S|\alpha_S[F]|^2,
\label{eq:total_walsh_norm}
\end{equation}
and the spectral weight above a truncation weight $D$ is
\begin{equation}
W_{>D}[F]
=
\sum_{w=D+1}^{n_q}W_w[F].
\label{eq:high_weight_tail}
\end{equation}
We use the normalized quantity
\begin{equation}
\varepsilon_D[F]
=
\frac{W_{>D}[F]}{\mathcal{N}_F}
\label{eq:normalized_compressibility_error}
\end{equation}
as a measure of Walsh compressibility. A small
$\varepsilon_D[F]$ indicates that the dominant part of the operator is
contained in Pauli sectors of weight at most $D$.

Since there are $\binom{n_q}{w}$ Pauli strings of weight $w$, the complexity of the Walsh representation is determined not
only by the number of available strings, but also by the magnitude and
distribution of their coefficients.

\subsection{General truncation bound}
\label{subsec:general_truncation_bound}

Consider first a polynomial function of degree $D$,
\begin{equation}
P_D(p)
=
\sum_{a=0}^{D}a_a p^a.
\label{eq:polynomial_D}
\end{equation}
The support result of Sec.~\ref{sec:exact_spectrum} implies
\begin{equation}
\deg_W\!\left[P_D(\hat p)\right]\leq D.
\end{equation}
Hence, a polynomial of degree $D$ contains no Walsh--Pauli components
of weight larger than $D$.

For a general analytic function, write
\begin{equation}
F(p)
=
P_D(p)+R_D(p),
\label{eq:F_poly_remainder}
\end{equation}
where
\begin{equation}
R_D(p)
=
\sum_{a=D+1}^{\infty}a_a p^a.
\label{eq:R_D}
\end{equation}
Since all terms in $P_D(\hat p)$ have Pauli weight at most $D$, the
high-weight spectrum of $F$ originates entirely from the remainder:
\begin{equation}
W_{>D}[F]
=
W_{>D}[R_D].
\label{eq:tail_remainder}
\end{equation}

Parseval's identity gives
\begin{equation}
\sum_S|\alpha_S[R_D]|^2
=
\frac{1}{N}
\sum_{\ell=0}^{N-1}
|R_D(p_\ell)|^2.
\end{equation}
Since $W_{>D}[R_D]$ is only part of the total Walsh norm,
\begin{equation}
W_{>D}[F]
\leq
\frac{1}{N}
\sum_{\ell=0}^{N-1}
|R_D(p_\ell)|^2.
\label{eq:tail_parseval}
\end{equation}
If the momentum lattice lies within
$|p|\leq p_{\max}$, this yields
\begin{equation}
W_{>D}[F]
\leq
\max_{|p|\leq p_{\max}}
|R_D(p)|^2 .
\label{eq:global_tail_bound}
\end{equation}

This establishes a direct connection between the convergence of the
analytic function of the momentum operator and the suppression of high-weight
Pauli sectors. A uniform bound on the remainder $R_D$ therefore gives
a rigorous bound on the discarded Walsh spectral weight.

The bound is deliberately conservative. It does not use the detailed
distribution of coefficients within a given shell. The exact
coefficient structure derived in Sec.~\ref{sec:exact_spectrum} allows
this additional information to be retained.

\subsection{Shell-resolved bound}
\label{subsec:shell_resolved_bound}

For a fixed Pauli weight $w$, the support and parity results give
\begin{equation}
\alpha_S[F]
=
\sum_{\substack{a\geq w\\a\equiv w\,(\mathrm{mod}\,2)}}
a_a c_{a,S},
\qquad
|S|=w.
\label{eq:fixed_shell_alpha}
\end{equation}
Using
\begin{equation}
|c_{a,S}|
\leq
p_{\max}^{\,a},
\end{equation}
we obtain
\begin{equation}
|\alpha_S[F]|
\leq
\sum_{\substack{a\geq w\\a\equiv w\,(\mathrm{mod}\,2)}}
|a_a|p_{\max}^{\,a}.
\label{eq:fixed_shell_bound}
\end{equation}
Since there are $\binom{n_q}{w}$ subsets of weight $w$,
\begin{equation}
W_w[F]
\leq
\binom{n_q}{w}
\left[
\sum_{\substack{a\geq w\\a\equiv w\,(\mathrm{mod}\,2)}}
|a_a|p_{\max}^{\,a}
\right]^2 .
\label{eq:shell_bound}
\end{equation}

This bound makes the origin of high-weight suppression transparent.
A weight-$w$ sector receives contributions only from Taylor orders
$a\geq w$ with the same parity as $w$. Thus, high Pauli weight probes
progressively higher orders in the analytic expansion of $F$.

The preceding estimate treats all coefficients within a shell
uniformly. The leading coefficient derived in
Sec.~\ref{sec:exact_spectrum} retains the additional intra-shell
structure. For $|S|=w$,
\begin{equation}
\alpha_S[F]
=
a_w c_{w,S}
+
R_{w,S},
\label{eq:leading_shell_decomposition}
\end{equation}
where
\begin{equation}
R_{w,S}
=
\sum_{\substack{a>w\\a\equiv w\,(\mathrm{mod}\,2)}}
a_a c_{a,S}.
\label{eq:higher_shell_terms}
\end{equation}
Since $c_{w,S}$ satisfies \eqref{eq:leading_coefficient} we obtain
\begin{equation}
|\alpha_S[F]|
\leq
|a_w|\,w!\prod_{r\in S}|c_r|
+
\sum_{\substack{a>w\\a\equiv w\,(\mathrm{mod}\,2)}}
|a_a|p_{\max}^{\,a}.
\label{eq:improved_coefficient_bound}
\end{equation}
Consequently,
\begin{equation}
W_w[F]
\leq
\sum_{|S|=w}
\left[
|a_w|\,w!\prod_{r\in S}|c_r|
+
\sum_{\substack{a>w\\a\equiv w\,(\mathrm{mod}\,2)}}
|a_a|p_{\max}^{\,a}
\right]^2 .
\label{eq:improved_shell_bound}
\end{equation}

For the JLP coefficients,
\begin{equation}
|c_r|\propto 2^r,
\end{equation}
and therefore the leading intra-shell contribution scales as
\begin{equation}
\prod_{r\in S}|c_r|
\propto
2^{\sum_{r\in S}r}.
\label{eq:JLP_intrashell}
\end{equation}
The Walsh spectrum is consequently nonuniform within a fixed
Pauli-weight shell. The analytic structure of $F$ controls the
suppression between different shells, while the binary structure of
the JLP momentum operator determines how the weight is distributed
within an individual shell.

When the leading allowed Taylor contribution dominates the higher-order
terms, the shell weight takes the approximate form
\begin{equation}
W_w[F]
\simeq
|a_w|^2(w!)^2
\sum_{|S|=w}
\prod_{r\in S}|c_r|^2 .
\label{eq:leading_shell_weight}
\end{equation}
For
\begin{equation}
c_r=-\kappa 2^r,
\qquad
\kappa=\frac{\pi}{N\delta x},
\end{equation}
this becomes
\begin{equation}
W_w[F]
\simeq
|a_w|^2(w!)^2\kappa^{2w}
\sum_{|S|=w}
2^{2\sum_{r\in S}r}.
\label{eq:JLP_shell_weight}
\end{equation}
Introducing the elementary symmetric polynomial
\begin{equation}
e_w(1,4,\ldots,4^{n_q-1})
=
\sum_{|S|=w}
4^{\sum_{r\in S}r},
\end{equation}
we obtain
\begin{equation}
W_w[F]
\simeq
|a_w|^2(w!)^2\kappa^{2w}
e_w(1,4,\ldots,4^{n_q-1}).
\label{eq:elementary_shell}
\end{equation}

Equation~\eqref{eq:elementary_shell} isolates the two ingredients
controlling the leading shell weight: the Taylor coefficient $a_w$ of
the analytic function and the binary hierarchy of the JLP momentum
operator. This separation will be used in the following section to
analyze a concrete deformed momentum relation.

The preceding analysis is independent of the detailed choice of the
analytic function $F$. We now apply it to generalized
uncertainty-principle (GUP) kinematics, which provides a physically
motivated non-polynomial momentum function. Its expansion contains
successively higher odd powers of momentum, making it a useful example
for studying how analytic structure is translated into Walsh-shell
structure and compressibility.

\section{Application to GUP Kinematics}
\label{sec:GUP_application}

The general framework developed above applies to any analytic function
of the JLP momentum operator. We now consider quadratic generalized
uncertainty-principle (GUP) kinematics as a concrete physical
application. Generalized uncertainty-principle models provide effective
descriptions of possible modifications of canonical quantum kinematics
and can lead to nonlinear relations between momentum and wave number
\cite{Banerjee:2010sd,KempfManganoMann1995,Kempf1995,Hossenfelder2013, Hossenfelder:2006cw}.

For the quadratic GUP, the modified commutation relation may be written
as
\begin{equation}
[\hat x,\hat p]
=
i\hbar\left(1+\beta\hat p^{\,2}\right),
\label{eq:GUP_commutator}
\end{equation}
where $\beta$ characterizes the deformation from the canonical
Heisenberg algebra. A corresponding wave-number variable $k$ satisfies
\begin{equation}
\frac{dk}{dp}
=
\frac{1}{1+\beta p^2}.
\label{eq:GUP_dkdp}
\end{equation}
Choosing $k(0)=0$ gives
\begin{equation}
k_\beta(p)
=
\frac{1}{\sqrt{\beta}}
\arctan(\sqrt{\beta}\,p).
\label{eq:GUP_spectral_function}
\end{equation}

On the finite JLP register, the corresponding operator is therefore
\begin{equation}
\hat k_\beta
=
k_\beta(\hat p).
\label{eq:GUP_operator}
\end{equation}
No new Walsh basis is required: $k_\beta(\hat p)$ is another analytic
function of the momentum operator and is consequently covered by the
generating-function construction of Sec.~\ref{sec:exact_spectrum}.

For a fixed $n_q$-qubit register, the Walsh--Pauli expansion is finite,
with maximum Pauli weight $n_q$. In contrast to a polynomial of fixed
degree, the non-polynomial function $k_\beta(p)$ has no
register-independent finite-degree cutoff in Pauli weight. As the
register is enlarged, progressively higher odd-weight sectors can
therefore become available. The parameter $\beta$ controls the relative
importance of the higher-order terms, providing a simple setting in
which to examine their contribution to the Walsh spectrum and its
compressibility.

\subsection{GUP-induced Walsh sectors}
\label{subsec:GUP_Walsh_structure}

Expanding Eq.~\eqref{eq:GUP_spectral_function} about $p=0$ gives
\begin{equation}
k_\beta(p)
=
p-\frac{\beta}{3}p^3
+\frac{\beta^2}{5}p^5
-\frac{\beta^3}{7}p^7+\cdots ,
\label{eq:GUP_Taylor}
\end{equation}
or, equivalently,
\begin{equation}
k_\beta(p)
=
\sum_{j=0}^{\infty}
\frac{(-1)^j}{2j+1}
\beta^j p^{2j+1}.
\label{eq:GUP_general_series}
\end{equation}

The odd parity of $k_\beta(p)$ immediately activates the parity
selection rule derived in Sec.~\ref{sec:exact_spectrum}:
\begin{equation}
W_w[k_\beta]=0,
\qquad
w\ {\rm even}.
\label{eq:GUP_odd_shells}
\end{equation}
Only odd Pauli-weight sectors can therefore contribute.

The structure within the odd sectors follows from the support theorem.
The term $p^{2j+1}$ can populate only the weights
\begin{equation}
w=1,3,\ldots,2j+1.
\end{equation}
Thus the leading term $p$ contributes only to weight $1$, while the
first deformation correction,
\begin{equation}
-\frac{\beta}{3}\hat p^3,
\end{equation}
can generate weight $3$ in addition to weight $1$. The next correction,
\begin{equation}
\frac{\beta^2}{5}\hat p^5,
\end{equation}
can additionally populate weight $5$, and higher-order terms extend
this pattern to progressively larger odd weights.

The GUP expansion therefore provides a direct illustration of the
hierarchy
\begin{equation}
1
\longrightarrow
3
\longrightarrow
5
\longrightarrow
7
\longrightarrow\cdots .
\label{eq:GUP_weight_sequence}
\end{equation}
The appearance of a higher Pauli-weight sector is tied directly to a
higher-order term in the analytic expansion of the deformed momentum
relation. In this sense, the Walsh representation provides an
operator-level view of how nonlinear analytic structure is distributed
across Pauli-weight sectors.
\subsection{Deformation-induced shell hierarchy}
\label{subsec:GUP_shell_hierarchy}

The relevant dimensionless measure of the deformation over the finite
momentum range is
\begin{equation}
\epsilon
=
\beta p_{\max}^{\,2},
\label{eq:GUP_epsilon}
\end{equation}
where $p_{\max}$ is the largest momentum represented on the finite
register. The ratio between successive terms in
Eq.~\eqref{eq:GUP_Taylor} is parametrically controlled by
\begin{equation}
\frac{
\beta^{j+1}p_{\max}^{2j+3}
}{
\beta^j p_{\max}^{2j+1}
}
\sim
\beta p_{\max}^2
=
\epsilon.
\label{eq:GUP_term_ratio}
\end{equation}
Thus,
\begin{equation}
\epsilon\ll1
\label{eq:weak_GUP}
\end{equation}
defines the weak-deformation regime in which the higher-order
corrections are successively suppressed.

Let $\alpha_S^{(\beta)}$ denote the Walsh coefficient of
$k_\beta(\hat p)$. Using Eq.~\eqref{eq:GUP_general_series} together
with the coefficient-extraction result of
Sec.~\ref{sec:exact_spectrum},
\begin{equation}
\alpha_S^{(\beta)}
=
\sum_{j=0}^{\infty}
\frac{(-1)^j}{2j+1}
\beta^j c_{2j+1,S}.
\label{eq:GUP_alpha}
\end{equation}

Consider an odd-weight subset with $|S|=w$. The support theorem implies
that the first possible contribution occurs at the power $p^w$.
Consequently, the leading contribution to the coefficient is
\begin{equation}
\alpha_S^{(\beta)}
=
\frac{(-1)^{(w-1)/2}}{w}
\beta^{(w-1)/2}
c_{w,S}
+
\mathcal{O}
\left(
\beta^{(w+1)/2}
\right),
\quad
w\ {\rm odd}.
\label{eq:GUP_leading_alpha}
\end{equation}
Using the leading monomial coefficient from
Eq.~\eqref{eq:leading_coefficient}, we obtain
\begin{equation}
\alpha_S^{(\beta)}
=
(-1)^{(w-1)/2}
(w-1)!
\beta^{(w-1)/2}
\prod_{r\in S}c_r
+
\mathcal{O}
\left(
\beta^{(w+1)/2}
\right).
\label{eq:GUP_leading_alpha_explicit}
\end{equation}

At fixed register parameters, the leading contribution to an
individual weight-$w$ coefficient therefore carries the characteristic
deformation dependence
\begin{equation}
|\alpha_S^{(\beta)}|
\propto
\beta^{(w-1)/2}.
\label{eq:GUP_alpha_scaling}
\end{equation}
The corresponding leading shell weight is
\begin{equation}
W_w[k_\beta]
\simeq
\frac{\beta^{w-1}}{w^2}
\sum_{|S|=w}|c_{w,S}|^2,
\qquad
w\ {\rm odd},
\label{eq:GUP_shell_weight}
\end{equation}
or, using
$c_{w,S}=w!\prod_{r\in S}c_r$,
\begin{equation}
W_w[k_\beta]
\simeq
\beta^{w-1}(w-1)!^2
\sum_{|S|=w}
\prod_{r\in S}|c_r|^2.
\label{eq:GUP_shell_weight_explicit}
\end{equation}

For the JLP coefficients Eq.~\eqref{eq:c_r}, we can define $\kappa$ such that
\begin{equation}
c_r=-\kappa 2^r,
\qquad
\kappa=\frac{\pi}{N\delta x}.
\end{equation}
Then Eq.~\eqref{eq:GUP_shell_weight_explicit} becomes
\begin{align}\label{eq:GUP_JLP_shell}
W_w[k_\beta]
&\simeq
\beta^{w-1}(w-1)!^2
\kappa^{2w}
\sum_{|S|=w}
4^{\sum_{r\in S}r}\\
&\equiv W^{LO}_w[k_\beta],
\end{align}
where we have defined the leading-shell approximation \eqref{eq:GUP_JLP_shell} to be $W^{LO}_w[k_\beta]$.

\begin{figure}[t]
    \centering
    \includegraphics[width=\columnwidth]{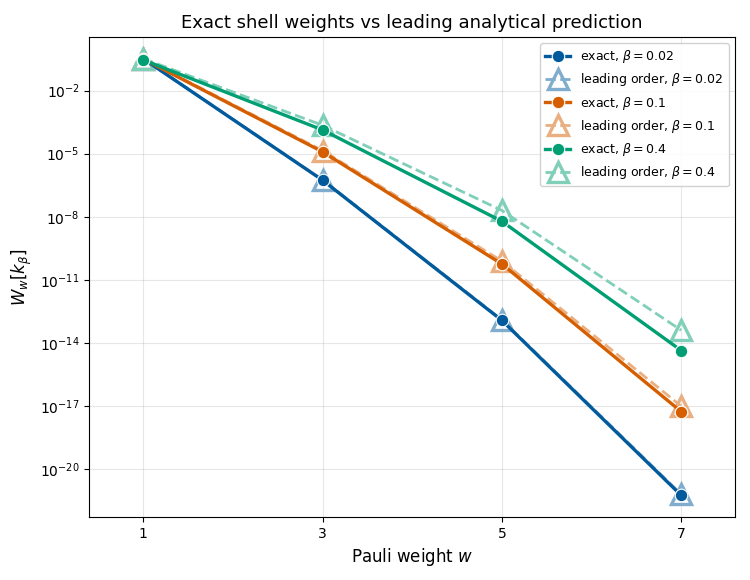}
\caption{
    Exact Walsh shell weights of the quadratic-GUP operator compared
    with the leading-order analytical prediction
    Eq.~\eqref{eq:GUP_JLP_shell} for $n_q=8$
    and $p_{\max}\approx0.996$.
    The agreement is excellent in the weak-deformation regime, while
    deviations appear as higher Taylor orders become appreciable.}
    \label{fig:GUP_exact_vs_leading}
\end{figure}

Figure~\ref{fig:GUP_exact_vs_leading} demonstrates that the
leading-order expression Eq.~\eqref{eq:GUP_JLP_shell}
accurately reproduces the exact finite-lattice Walsh spectrum in the
weak-deformation regime.

The numerical spectrum validates the leading-order shell prediction over the tested parameter range, with near-perfect agreement in the weak-deformation regime and controlled deviations as the deformation becomes stronger.

To quantify the accuracy of the leading-shell approximation more clearly, we define
\begin{equation}
\mathcal{R}_w(\beta)
=
\frac{
W_w^{\rm exact}[k_\beta]
}{
W_w^{\rm LO}[k_\beta]
}.
\label{eq:GUP_LO_ratio}
\end{equation}
The leading-order approximation corresponds to
$\mathcal{R}_w\simeq1$.

\begin{figure}[t]
    \centering
    \includegraphics[width=\columnwidth]{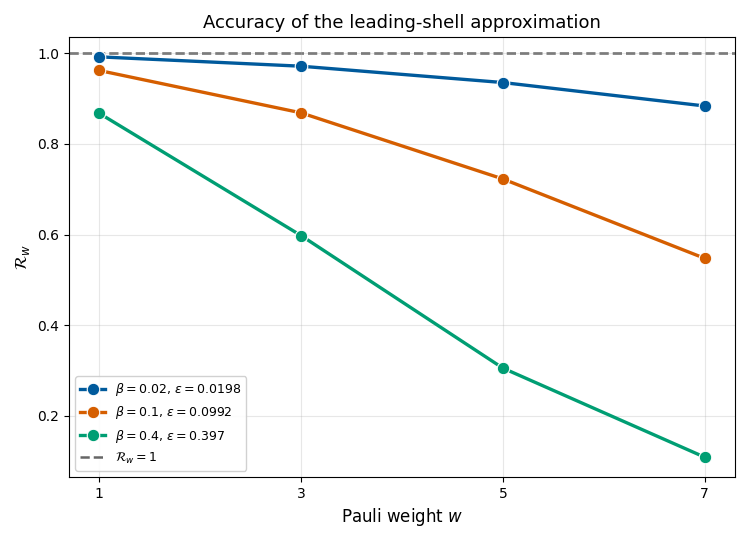}
    \caption{
    Ratio
    $\mathcal{R}_w=W_w^{\rm exact}/W_w^{\rm LO}$ defined in
    Eq.~\eqref{eq:GUP_LO_ratio}. Values near unity indicate dominance
    of the leading Taylor contribution to the corresponding
    Pauli-weight shell.
    }
    \label{fig:GUP_leading_ratio}
\end{figure}

Figure~\ref{fig:GUP_leading_ratio} quantifies the accuracy of the
leading-shell approximation. In the weak-deformation regime the ratio
remains close to unity, while increasing deviations at larger
$\epsilon$ signal the growing contribution of higher Taylor orders
within the same Pauli-weight sector.

Crucially, we see once again through Equation~\eqref{eq:GUP_leading_alpha} the two ingredients
of the shell hierarchy are seperated. The Taylor coefficients of the GUP function
determine the dependence on the deformation parameter, while the
binary coefficients $c_r$ determine the distribution of weight within
each Pauli shell.

\subsection{Walsh compressibility of the GUP operator}
\label{subsec:GUP_compressibility}

The shell hierarchy directly determines the degree to which the GUP
operator can be represented using low-weight Pauli strings. Define the
cumulative spectral weight retained up to Pauli weight $D$ as
\begin{equation}
C(D)
=
\frac{
\displaystyle\sum_{w=0}^{D}W_w[k_\beta]
}{
\displaystyle\sum_{w=0}^{n_q}W_w[k_\beta]
},
\label{eq:GUP_cumulative}
\end{equation}
and the corresponding discarded fraction as
\begin{equation}
\varepsilon_D[k_\beta]
=
1-C(D)
=
\frac{
\displaystyle\sum_{w>D}W_w[k_\beta]
}{
\displaystyle\sum_{w=0}^{n_q}W_w[k_\beta]
}.
\label{eq:GUP_discarded}
\end{equation}

For the odd GUP spectral function, only odd-weight shells contribute.
In the weak-deformation regime $\epsilon\ll1$, the higher-order
contributions are suppressed and the cumulative weight is expected to
saturate after a small number of low-weight sectors. As $\epsilon$
increases, progressively higher odd shells become relevant and a
larger Pauli weight is required to achieve the same truncation
accuracy.

\begin{figure}[t]
    \centering
    \includegraphics[width=\columnwidth]{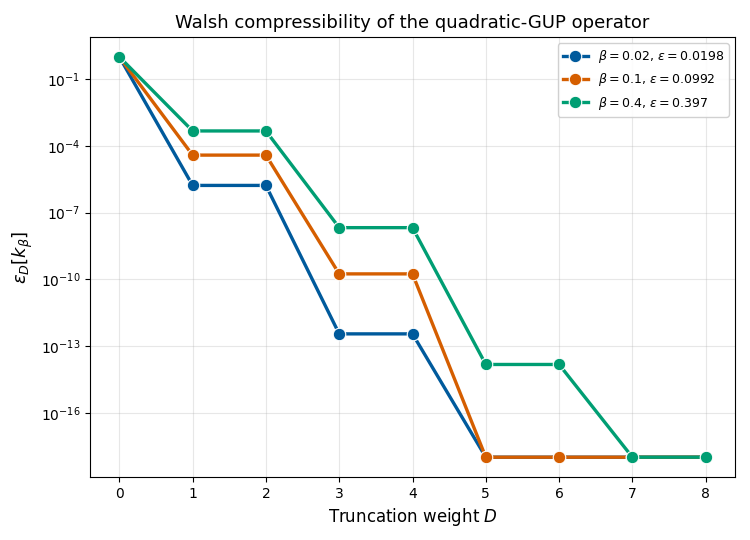}
    \caption{
    Discarded Walsh spectral fraction
    $\varepsilon_D[k_\beta]=1-C(D)$,
    defined in Eqs.~\eqref{eq:GUP_cumulative}
    and~\eqref{eq:GUP_discarded}, as a function of the maximum retained
    Pauli weight $D$. Increasing the deformation transfers spectral
    weight into higher-weight sectors and therefore increases the
    Pauli locality required at fixed accuracy.
    }
    \label{fig:GUP_compressibility}
\end{figure}

Figure~\ref{fig:GUP_compressibility} shows that weak GUP deformations
are strongly compressible in Pauli weight, while increasing the
deformation populates progressively higher-weight sectors.

For a prescribed tolerance $\varepsilon_{\rm tol}$, we define the
required Walsh weight $D_\ast$ by
\begin{equation}
D_\ast
=
\min
\left\{
D:\,
\varepsilon_D[k_\beta]
\leq
\varepsilon_{\rm tol}
\right\}.
\label{eq:GUP_required_weight}
\end{equation}
This provides a direct operator-level measure of the locality required
to represent the GUP-modified momentum to a specified accuracy.

The analytical predictions can be tested against the exact Walsh
spectrum obtained from the finite JLP lattice. For fixed $n_q$ and
$p_{\max}$, varying $\beta$, or equivalently
$\epsilon=\beta p_{\max}^2$, allows one to track the redistribution of
spectral weight among the allowed odd shells. In particular, the
calculation tests three predictions of the analytical framework:
\begin{enumerate}
\item the vanishing of all even-weight shells;
\item the progressive population of higher odd shells with increasing
deformation;
\item the increase of the required truncation weight $D_\ast$ as the
deformation becomes stronger.
\end{enumerate}

The GUP example therefore provides a concrete physical realization of
the general connection established in this work between the analytic
structure of a diagonal operator and the locality and compressibility
of its Walsh--Pauli representation.

\section{Universality and Model Dependence of the Walsh Spectrum}
\label{sec:other_GUPs}

The quadratic-GUP example of Sec.~\ref{sec:GUP_application} illustrates
how the analytic structure of a nonlinear momentum function is reflected
in its Walsh spectrum. In particular, odd parity restricts the spectrum
to odd Pauli-weight sectors, while higher-order terms in the analytic
expansion can generate progressively higher Walsh shells. These
features are not specific to the quadratic GUP. More generally, the
selection rules are determined by the analytic and symmetry properties
of the spectral function, whereas the detailed distribution of spectral
weight depends on its coefficients.

To make this distinction explicit, we consider a representative family
of analytic deformations defined on the same JLP momentum lattice. The
purpose is not to introduce separate Walsh constructions for different
models, but to apply the common generating-function framework to
different spectral functions and thereby identify which properties are
fixed by the encoding and which depend on the deformation.

\subsection{A family of analytic deformations}
\label{subsec:GUP_family}

Consider a class of deformed commutation relations \cite{PhysRevD.52.1108,Bosso:2023aht} of the form 
\begin{equation}
[\hat x,\hat p]
=
i\hbar\,g(\hat p^{\,2}),
\label{eq:general_GUP_commutator}
\end{equation}
for which the corresponding wave-number function \cite{Bosso:2020aqm} can be written as
\begin{equation}
k(p)
=
\int_0^p
\frac{dp'}{g(p'^{\,2})}.
\label{eq:general_spectral_function}
\end{equation}
After discretization, each spectral function $k(p)$ is treated using
the same Walsh--Pauli construction developed in
Sec.~\ref{sec:exact_spectrum}.

We consider the representative spectral functions listed in
Table~\ref{tab:GUP_models}. The quadratic-GUP model discussed in
Sec.~\ref{sec:GUP_application} is included as the reference case.

\begin{table}[t]
\centering
\caption{
Representative analytic deformations considered in this section,
motivated by the examples discussed in Ref.~\cite{Bosso2023minimal}.
All listed models are odd functions of momentum. The final column gives
the first nonlinear contribution beyond the canonical term.
}
\label{tab:GUP_models}
\small
\renewcommand{\arraystretch}{1.25}
\begin{tabular}{lcc}
\hline
Model & $k(p)$ & Leading nonlinear term \\
\hline
Quadratic GUP
&
$\displaystyle
\frac{1}{\sqrt{\beta}}
\arctan(\sqrt{\beta}\,p)$
&
$\displaystyle
-\frac{\beta}{3}p^3$
\\[6pt]

Arctanh
&
$\displaystyle
\frac{1}{\sqrt{\beta}}
\operatorname{arctanh}(\sqrt{\beta}\,p)$
&
$\displaystyle
+\frac{\beta}{3}p^3$
\\[6pt]

Error function
&
$\displaystyle
\frac{\sqrt{\pi}}{2\sqrt{\beta}}
\operatorname{erf}(\sqrt{\beta}\,p)$
&
$\displaystyle
-\frac{\beta}{3}p^3$
\\[6pt]

Polynomial
&
$\displaystyle
p-\frac{\beta}{3}p^3$
&
$\displaystyle
-\frac{\beta}{3}p^3$
\\[6pt]

Arcsine
&
$\displaystyle
\frac{1}{\sqrt{\beta}}
\arcsin(\sqrt{\beta}\,p)$
&
$\displaystyle
+\frac{\beta}{6}p^3$
\\
\hline
\end{tabular}
\end{table}

All models in Table~\ref{tab:GUP_models} are odd functions of
momentum and therefore, by the parity selection rule of
Sec.~\ref{sec:exact_spectrum}, have support only on odd Pauli-weight
sectors. Thus, the set of allowed shells is universal across this
family. The detailed shell weights, however, depend on the magnitudes
and relative signs of the higher-order coefficients in the corresponding
analytic expansions.

The polynomial model provides a simple limiting case. Since its
expansion terminates at cubic order, the support theorem implies that
it has no contribution from Pauli weights above three. The non-polynomial
models, by contrast, contain higher powers of momentum and can therefore
populate progressively higher odd-weight sectors. Their Walsh spectra
can consequently exhibit different degrees of spectral concentration
and compressibility even though they obey the same parity selection
rule.

The comparison therefore separates two aspects of the digital
representation. The allowed Pauli-weight sectors follow from the
analytic structure and symmetry of the spectral function, while the
distribution of spectral weight within those sectors remains
model-dependent.

\subsection{Universal lattice structure and model dependence}
\label{subsec:universal_lattice}

Let a general analytic spectral function be written as
\begin{equation}
k(p)
=
\sum_{m=0}^{\infty}a_m p^m.
\label{eq:general_k_series}
\end{equation}
The corresponding Walsh coefficients are
\begin{equation}
\alpha_S[k]
=
\sum_{m=0}^{\infty}a_m c_{m,S},
\label{eq:general_alpha_family}
\end{equation}
where $c_{m,S}$ denotes the Walsh coefficient of $\hat p^m$ obtained
from the generating-function construction.

Equation~\eqref{eq:general_alpha_family} separates the universal
JLP structure from the model-dependent deformation. For a fixed
momentum lattice, the coefficients $c_{m,S}$ are determined entirely
by the Walsh--Pauli decomposition of the momentum powers $\hat p^m$
and are therefore common to all spectral functions considered here.
The choice of deformation enters only through the Taylor coefficients
$a_m$ of $k(p)$.

Consequently, once the monomial coefficients $c_{m,S}$ have been
determined for a given register, the Walsh spectrum of any analytic
deformation can be constructed by combining this same set of
coefficients with the corresponding $a_m$. Changing the deformation
therefore changes how the universal momentum-power contributions are
combined, but does not require a new construction of the underlying
Walsh structure.

\subsection{Finite and infinite Walsh support on arbitrary qubits}
\label{subsec:model_support}

The distinction between polynomial and non-polynomial deformations is
particularly transparent at the level of Walsh support. For the
polynomial model,
\begin{equation}
k_{\rm poly}(p)
=
p-\frac{\beta p^3}{3},
\end{equation}
the expansion terminates at cubic order. The support theorem therefore
implies
\begin{equation}
W_w[k_{\rm poly}]=0,
\qquad
w>3.
\label{eq:poly_finite_support}
\end{equation}
The polynomial deformation therefore possesses an exact
register-independent cutoff at weight $w=3$. More generally, a
polynomial of degree $D$ has no Walsh--Pauli support above $w=D$,
irrespective of the register size.

The remaining analytic models contain infinitely many momentum powers
and therefore possess no analogous register-independent weight cutoff.
For any fixed $n_q$, their Walsh spectra remain finite with
$w\leq n_q$, but progressively higher odd-weight sectors become
available as the register is enlarged. Their practical compressibility
is therefore determined not by an exact support cutoff, but by the
suppression of the high-weight spectral tail.

\subsection{Interference between momentum orders}
\label{subsec:sign_cancellation}

The distribution of spectral weight cannot, in general, be inferred
from the magnitudes of the Taylor coefficients alone. For a fixed
subset $S$, the Walsh coefficient is the coherent sum defined via \eqref{eq:general_alpha_family}.

Because the shell weight is constructed from the squared magnitude of the full coefficient  $\alpha_S[k]$, it is not, in general, equal to a simple sum of the shell weights associated with the individual monomials. Contributions from different momentum orders can instead reinforce or partially cancel one another.

This effect is relevant when comparing the models in
Table~\ref{tab:GUP_models}. Even when two deformations have comparable
magnitudes of higher-order coefficients, their Walsh spectra can differ
because those coefficients enter with different relative signs and
weights. The Walsh spectrum therefore reflects the coherent structure
of the analytic deformation as well as the magnitude of its individual
Taylor terms.

\subsection{Walsh-shell comparison}
\label{subsec:family_shell_comparison}

We now compare the Walsh-shell distributions of the representative
non-polynomial deformations in Table~\ref{tab:GUP_models}, evaluated
on the same JLP momentum lattice. For fixed register size and momentum
cutoff, the deformation strength is characterized by the dimensionless
parameter
\begin{equation}\label{eps}
\epsilon=\beta p_{\max}^2,
\end{equation}
introduced in Eq.~\eqref{eq:GUP_epsilon}. Using the same value of
$\epsilon$ therefore provides a common scale for comparing the
different spectral functions.

The comparison separates the universal structure associated with the
JLP representation from the model-dependent structure of the
deformation. Since all spectral functions considered here are odd,
their Walsh spectra are restricted to odd Pauli-weight sectors. The
JLP lattice fixes the monomial coefficients $c_{m,S}$, while the
Taylor coefficients of each spectral function determine how the
different momentum orders combine within these allowed sectors.

\begin{figure}[t]
    \centering
    \includegraphics[width=\columnwidth]{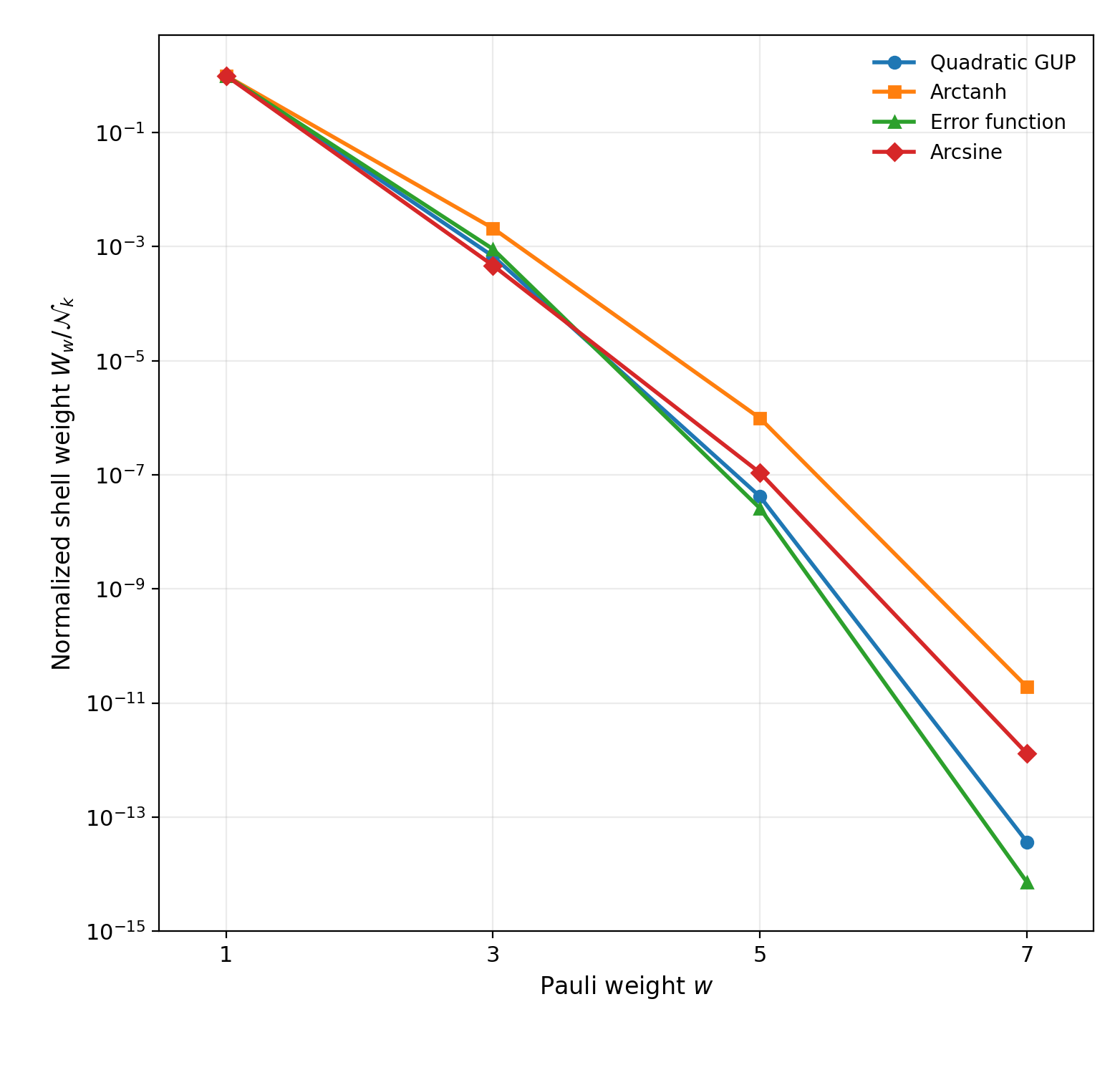}
    \caption{
    Normalized Walsh shell weights for the representative
    non-polynomial analytic deformations of
    Table~\ref{tab:GUP_models}, evaluated on the same JLP momentum
    lattice at fixed deformation strength
    $\epsilon=\beta p_{\max}^2$.
    All models populate only odd Pauli-weight sectors, reflecting
    their common parity, while the relative distribution of spectral
    weight among the allowed shells depends on the analytic form of
    the deformation. The values shown are $n_q=8$, $\epsilon=0.5$ and $p_{\max}=1$.
    }
    \label{fig:general_deformation_shells}
\end{figure}

For each model, the Walsh coefficients are evaluated on the finite
JLP register and grouped according to Pauli weight. To compare their
relative shell distributions independently of the overall operator
norm, we consider the normalized shell weights
$W_w[k]/\mathcal{N}_k$, where $\mathcal{N}_k$ is the total Walsh norm.
The resulting distributions are shown in
Fig.~\ref{fig:general_deformation_shells}.

Figure~\ref{fig:general_deformation_shells} makes the distinction
between universal and model-dependent features explicit. The common
restriction to odd Pauli weights follows solely from the parity of the
spectral functions and is therefore independent of the detailed form
of the deformation. In contrast, the relative population of the
allowed shells differs between models, reflecting the different
Taylor coefficients and their coherent combination within each Walsh
sector. The higher-weight contributions are progressively suppressed,
but the rate and detailed pattern of this suppression are
model dependent.
\subsection{Relative Walsh compressibility}

\label{subsec:relative_compressibility}

The compressibility measures introduced for the quadratic-GUP operator
in Sec.~\ref{subsec:GUP_compressibility} apply equally to a general
spectral function $k(p)$. We therefore use the cumulative retained
weight $C(D)$, discarded fraction $\varepsilon_D[k]$, and minimum
required Pauli weight $D_{\rm min}$ defined analogously to
Eqs.~\eqref{eq:GUP_cumulative}--\eqref{eq:GUP_required_weight}.
These quantities provide a common measure for comparing the Walsh
compressibility of the different deformations.
For the polynomial deformation,
Eq.~\eqref{eq:poly_finite_support} implies
\begin{equation}
D_{\rm min}\leq3
\end{equation}
for an exact representation. The analytic deformations have no
corresponding finite-support bound, but their effective support can
remain small when their high-weight contributions are sufficiently
suppressed.

The quantity $D_{\rm min}$ therefore provides a common measure with
which the digital complexity of different deformations can be
compared. It incorporates both the deformation-dependent analytic
structure and the universal Walsh structure of the underlying momentum
lattice.

\subsection{Towards a Full Digital Realisation of GUP Kinematics}

The present analysis concerns the representation of the GUP spectral function $k_\beta(\hat p)$ on a fixed JLP momentum register and the Pauli resources associated with diagonal functions of this operator. It does not by itself constitute a complete finite-dimensional realization of GUP quantum mechanics. In particular, the nonlinear relation $k_\beta=k_\beta(p)$ modifies the relation between momentum and localization: applying $k_\beta$ to a uniformly discretized JLP momentum lattice produces a nonuniform wave-number lattice, for which the corresponding position--momentum transformation is not, in general, the standard discrete Fourier transform. Constructing a finite-dimensional transform consistent with the modified Hilbert-space measure and localization structure, and using it to simulate complete position--momentum dynamics, is a separate problem that we leave for future work. The present results provide one necessary ingredient for such a construction by determining analytically the Pauli representation and compressibility of the diagonal deformed-momentum operators.

\subsection{Implications for quantum-circuit implementation}
\label{subsec:family_circuit}

Because the Walsh--Pauli operators commute, the diagonal unitary
generated by $k(\hat p)$ factorizes as
\begin{equation}
e^{-itk(\hat p)}
=
\prod_S e^{-it\alpha_S Z_S}.
\label{eq:family_unitary}
\end{equation}
A Walsh truncation therefore corresponds directly to retaining only
those Pauli rotations whose coefficients contribute appreciably to the
desired accuracy.

The polynomial deformation has a register-independent upper bound on
its Walsh support. The analytic deformations have no analogous fixed
weight cutoff, although on every finite register their Walsh--Pauli
representations remain finite and are bounded by $w\leq n_q$. Their effective support can nevertheless remain concentrated at low
Pauli weight when the high-weight spectral tail is sufficiently
suppressed.

The comparison therefore establishes a direct connection between the
analytic structure of a deformation and the complexity of its digital
representation. The JLP momentum lattice fixes the available
Pauli structures, while the spectral function determines how the
spectral weight is distributed among them.

\section{Circuit Complexity of the Walsh Representation}
\label{sec:circuit_synthesis}

The analytical results obtained above determine not only the Walsh
coefficients of the diagonal operator but also their distribution
across Pauli-weight sectors. This structure has a direct consequence
for quantum implementation. A sparse Walsh spectrum requires only a
small number of Pauli-string rotations, while concentration in
low-weight sectors restricts the locality of the required operations.
The purpose of this section is to make this connection quantitative.

Using the Walsh--Pauli expansion of Eq.~\eqref{eq:F_p_pauli}, the
corresponding evolution factorizes exactly as
\begin{equation}
e^{-itF(\hat p)}
=
\prod_S e^{-it\alpha_S[F]Z_S},
\label{eq:exact_factorization}
\end{equation}
since all $Z_S$ commute. Thus each retained Walsh coefficient
corresponds to one Pauli-string rotation, and the factorization is
exact rather than Trotterized.

\subsection{From Walsh sparsity to circuit locality}
\label{subsec:walsh_to_circuit}

Let
\begin{equation}
N_w
=
\#\{S:\ |S|=w,\ \alpha_S[F]\neq0\}
\end{equation}
denote the number of nonzero Walsh coefficients in the weight-$w$
sector. The total number of Pauli rotations is then
\begin{equation}
N_{\rm rot}
=
\sum_{w=0}^{n_q}N_w.
\label{eq:number_rotations}
\end{equation}

The weight $w$ of a Walsh index specifies the number of qubits on
which the corresponding Pauli string acts. Hence, the largest
populated weight,
\begin{equation}
D_{\rm max}
=
\max\{w:N_w\neq0\},
\label{eq:Dmax}
\end{equation}
sets the maximum locality of the Walsh representation.

These two quantities capture different aspects of the circuit
complexity. The number $N_{\rm rot}$ measures the number of diagonal
Pauli rotations, whereas $D_{\rm max}$ measures their maximum
locality. The shell structure derived above therefore provides a
natural resource characterization of the operator before any explicit
circuit is constructed.

Weight alone does not determine the complete resource cost. As shown
by the intra-shell hierarchy above, coefficients within the same
weight sector can differ substantially. A practical synthesis may
therefore combine weight truncation with coefficient thresholding,
retaining only the individually significant strings within each
allowed shell.

\subsection{Weight truncation and approximate synthesis}\label{subsec:circuit_truncation}

The compressibility analysis provides a natural way to reduce these resources. The weight-truncated operator $F_D(\hat p)$ corresponds directly to a circuit retaining only Pauli rotations with $|S| \leq D$.

The corresponding discarded operator is
\begin{equation}
\Delta_D
=
F(\hat p)-F_D(\hat p)
=
\sum_{|S|>D}
\alpha_S[F]Z_S.
\label{eq:circuit_tail_operator}
\end{equation}

The discarded spectral content is precisely the high-weight tail
introduced in the compressibility analysis. Thus, the same quantity that measures Walsh compressibility also
determines the amount of operator structure omitted by the truncated
circuit.

For a prescribed relative spectral tolerance, the truncation weight is
therefore chosen using the quantity already introduced in
Eq.~\eqref{eq:normalized_compressibility_error},
\[
\varepsilon_D[F]\leq\varepsilon_{\rm tol}.
\]
The corresponding minimum value is precisely the required Walsh weight
$D_{\rm min}$ defined above.

This gives a direct trade-off:
reducing $D_{\rm min}$ therefore lowers the maximum locality of the retained
rotations at the cost of a larger discarded spectral tail.

\subsection{Elementary-gate cost}
\label{subsec:gate_cost}

A Pauli-string rotation
\begin{equation}
e^{-i\theta Z_S}
\end{equation}
can be implemented using a CNOT ladder, a single-qubit $R_z$
rotation, and the inverse CNOT ladder \cite{bullock2003smallercircuitsarbitrarynqubit}. For a weight-$w$ string, the
standard construction requires
\begin{equation}
2(w-1)
\end{equation}
CNOT gates and one single-qubit rotation.

For the weight-$D$ truncated representation, the corresponding
CNOT count is therefore bounded by
\begin{equation}
N_{\rm CNOT}
\leq
2\sum_{w=1}^{D}(w-1)N_w.
\label{eq:cnot_shell}
\end{equation}

Equation~\eqref{eq:cnot_shell} makes explicit how the Walsh shell
distribution enters the elementary-gate cost. A spectrum concentrated
in low-weight sectors requires only low-locality Pauli rotations,
whereas significant weight at large $w$ increases the entangling-gate
cost.

The precise gate count is implementation dependent, since different
connectivity constraints and circuit optimizations can reduce the
CNOT count. Equation~\eqref{eq:cnot_shell} should therefore be
understood as the cost of the standard Pauli-string decomposition
rather than as a circuit-optimality theorem.

\subsection{Quadratic GUP as a resource example}
\label{subsec:GUP_circuit_resources}

The quadratic-GUP deformation provides a direct physical example of
this general connection. In the canonical limit ($\beta \to 0$), the momentum
operator occupies only the weight-$1$ sector. Its diagonal evolution
can therefore be implemented using single-qubit $Z$ rotations.

The nonlinear GUP terms populate higher odd-weight sectors. In
particular, the cubic correction generates weight-$3$ Pauli strings,
while higher odd powers can generate weight-$5$, $7$, and higher
sectors. The Walsh-shell hierarchy derived in
Sec.~\ref{sec:GUP_application} therefore becomes a hierarchy of
circuit locality.

In the weak-deformation regime, the higher-weight sectors are
suppressed and the operator can be accurately represented by retaining
only the lowest odd-weight shells. Increasing the deformation
parameter transfers spectral weight towards higher shells and hence
increases the locality required for a faithful representation.

The dimensionless parameter \eqref{eps} therefore provides a convenient measure of the deformation-induced
growth of Walsh-based circuit complexity.

\subsection{Interpretation of the resource measure}
\label{subsec:resource_interpretation}

The preceding results establish a direct correspondence between the
analytical Walsh spectrum and the resources required for its quantum
implementation. The number of populated Walsh sectors determines how
many Pauli rotations are required, while their weight determines the
locality and, for a given decomposition, the associated entangling-gate
cost.

The main advantage of the analytical approach is that these resources
can be inferred from the structure of the spectral function without
first constructing the full numerical Walsh transform. The generating
function determines the allowed sectors and their coefficients, the
recursive construction provides an efficient means of obtaining them,
and the high-weight tail determines the accuracy--complexity trade-off
of a truncated circuit.

\begin{equation}
\begin{aligned}
F(\hat p)
&\longrightarrow \text{Walsh spectrum}\\
&\longrightarrow \text{Pauli weight}\\
&\longrightarrow \text{circuit resources}.
\end{aligned}
\label{eq:overall_resource_flow}
\end{equation}

The circuit analysis thus provides the operational interpretation of
the Walsh compressibility established above: analytical concentration
of the spectrum in low-weight sectors translates into a quantum
representation requiring only low-locality Pauli rotations.

\section{Discussion and Conclusions}
\label{sec:conclusion}

The central problem addressed in this work is how the analytical structure
of a momentum-dependent operator is reflected in its finite-qubit
Walsh--Pauli representation. The Walsh--Pauli correspondence and the use
of Walsh expansions for quantum-circuit synthesis are already well
established~\cite{Walsh1923,Welch2014,Hadfield:2018yak}. The specific
problem considered here is instead how the Walsh spectrum of an
analytically specified operator can be determined directly from the
operator and the underlying binary encoding, rather than by first
evaluating the function on the complete discrete lattice and performing
a numerical Walsh transform. This operator-level problem is the main
gap addressed in the present work.

We have developed a generating-function framework for determining the
Walsh--Pauli coefficients of powers of the Jordan--Lee--Preskill (JLP)
momentum operator. The construction exploits the fact that the JLP
momentum operator is a linear combination of mutually commuting
Pauli-$Z$ operators, allowing the coefficients to be obtained directly
from the operator structure. The construction extends to general
analytic functions of the momentum operator through their power-series
expansion. Thus, the Walsh spectrum is determined analytically from the
operator and its binary encoding, rather than being obtained only as the
output of a numerical transform over the complete momentum lattice. To
our knowledge, this operator-level generating-function construction for
the Walsh spectrum of powers of the JLP momentum operator has not been
presented previously.

An important outcome of this construction is that it exposes structural
properties of the Walsh spectrum that are not apparent from the
coefficients alone. The polynomial order determines the maximum
Pauli-weight sector that can be populated, while parity restricts the
allowed sectors further. At a fixed Pauli weight, the binary structure
of the JLP encoding produces a nonuniform hierarchy among the individual
Pauli strings. These results separate the encoding-dependent structure
of the spectrum from the model-dependent information contained in the
analytic function. The binary encoding determines the organization of
the momentum-power contributions, while the analytic coefficients
determine how those contributions are combined.

The generating-function formulation also admits a recursive construction
of the coefficients. By incorporating the qubit factors successively,
the recursion allows selected coefficients and Pauli-weight sectors to
be obtained without explicitly constructing the complete momentum-space
data set. This provides a practical analytical route to the part of the
spectrum most relevant for a truncated representation. The shell
decomposition developed in this work then organizes the spectrum
according to Pauli weight and provides a quantitative measure of the
spectral weight retained or discarded at a chosen truncation.

This has a direct relevance for digital quantum computation. Once the
Walsh--Pauli spectrum is known, the corresponding diagonal evolution can
be implemented through commuting Pauli-string rotations. The analytical
determination of the spectrum therefore makes it possible to identify
the relevant Pauli-weight sectors before circuit synthesis and to
estimate the locality and resource requirements of a truncated
implementation. In this sense, the present framework can serve as an
analytical preprocessing step for quantum simulation: rather than
constructing the full discrete operator and only afterwards determining
its Walsh structure, one can use the analytical structure to anticipate
which Pauli sectors are important and how much spectral weight is lost
under a chosen truncation.

The present focus on momentum-dependent operators is motivated by the
structure of the JLP discretization. In the JLP momentum register, the
momentum operator is represented as a linear combination of mutually
commuting Pauli-$Z$ operators. This makes analytic functions of momentum
particularly suitable for an operator-level Walsh--Pauli analysis and
allows the generating-function construction to be factorized over the
qubit register. The choice of momentum is therefore deliberate and is
dictated by the representation used in the present analysis, rather
than by an assumption that momentum-dependent operators are more
fundamental than position-dependent ones.

The position sector provides a natural complementary direction. A
corresponding analysis of position-dependent operators would require an
appropriate position-space representation and binary encoding. Treating
position and momentum simultaneously is a more substantial extension,
because they form a noncommuting operator pair. In the GUP setting, the
problem is further complicated by the modified relation between momentum
and wave number and by the associated nonuniform wave-number structure.
A complete finite-dimensional treatment of both position and momentum
would therefore require constructing the corresponding position
operator and the transformation relating the position and momentum
representations. This lies beyond the scope of the present work.

The quadratic generalized uncertainty principle (GUP) provides a
physically motivated application of the framework. Its nonlinear
momentum dependence contains successively higher odd powers, and the
JLP representation consequently maps these contributions into
successively higher odd Pauli-weight sectors. The GUP example therefore
illustrates how a modification of continuum quantum kinematics can be
translated into a concrete change in the structure of a finite-qubit
operator. In the weak-deformation regime, the higher-weight
contributions remain suppressed, whereas increasing deformation
redistributes spectral weight toward higher Pauli-weight sectors. The
result is a direct connection between the strength of the nonlinear
deformation, Walsh compressibility, and the locality required for its
digital representation.

The GUP analysis also demonstrates that the structural properties of the
spectrum should be distinguished from its detailed numerical
distribution. The allowed support and parity constraints arise from the
momentum-power structure and the binary encoding, whereas the relative
spectral weights depend on the particular analytic deformation.
Consequently, different nonlinear momentum relations can share the same
general selection rules while requiring substantially different
truncation levels and circuit resources. This distinction is important
for using Walsh spectra as a diagnostic of analytically structured
operators rather than simply as a collection of numerical coefficients.

The circuit analysis should therefore be understood as an operational
consequence of the analytical spectrum rather than as a complete quantum
simulation of GUP quantum mechanics. The present framework is restricted
to diagonal functions of a single JLP momentum register. It provides,
however, a concrete analytical preprocessing step for digital quantum
simulation: for a specified momentum-dependent operator, its Walsh
spectrum can be characterized before circuit synthesis, the dominant
Pauli-weight sectors can be identified, and the high-weight contribution
can be used to assess the cost and accuracy of a truncated
representation. The GUP example demonstrates how such information can
be related to a physically motivated deformation of quantum kinematics.

The framework is not restricted to the quadratic GUP. Because the
generating-function construction applies to general analytic functions
of the JLP momentum operator, it can be used to analyze other nonlinear
momentum relations and modified dispersion functions within the same
binary representation. This provides a common setting in which
different analytic deformations can be compared through their
Pauli-weight distributions, Walsh compressibility, and associated
digital implementation costs.

Several extensions follow naturally from the present results. A first
is to use the analytically predicted spectra in explicit quantum
simulations and test the relation between Walsh truncation, simulation
accuracy, and practical gate resources in the presence of hardware
connectivity and noise. A second is to extend the generating-function
and recursive constructions to multiple momentum modes and interacting
operators, where correlations between registers may introduce additional
spectral structure. A third is to investigate whether alternative
binary encodings can produce stronger concentration in low-weight
sectors and thereby reduce the resources required for digital
implementation. For GUP specifically, constructing the corresponding
finite-dimensional position--momentum realization would provide a
natural next step toward simulating the complete deformed kinematics.

In this sense, the present work provides an analytical connection between
continuum momentum dependence and its digital quantum representation.
The generating-function construction determines the Walsh--Pauli
spectrum directly from the operator structure, the resulting spectrum
quantifies its Pauli-weight distribution and compressibility, and these
properties provide a basis for estimating and optimizing the resources
needed for quantum simulation. The GUP application demonstrates how a
nonlinear modification of momentum dependence can be translated into
higher-locality structure in a finite-qubit representation. The
framework can therefore be used to guide the construction of
resource-efficient digital representations of analytically structured
momentum-dependent operators, while its extension to position-dependent
operators, complete position--momentum kinematics, multimode and
interacting systems, and ultimately experimentally implemented quantum
simulations provides natural directions for future work.

\section{Acknowledgements}

R.D.S. gratefully acknowledges the support of the National Institute for
Theoretical and Computational Sciences (NITheCS), which provided funding
that enabled this project to be carried out during the period between the
completion of his Honours degree and the beginning of his PhD studies at
Oxford. P.~N. acknowledges support from the RPF Postdoctoral Fellowship
Program. We are grateful to Maria Schuld for insightful comments and
constructive suggestions that helped improve the clarity and presentation
of the manuscript.

\bibliographystyle{apsrev4-2}
\bibliography{gw_phases}

@article{Hossenfelder:2006cw,
    author = "Hossenfelder, S.",
    title = "{Interpretation of quantum field theories with a minimal length scale}",
    eprint = "hep-th/0603032",
    archivePrefix = "arXiv",
    doi = "10.1103/PhysRevD.73.105013",
    journal = "Phys. Rev. D",
    volume = "73",
    pages = "105013",
    year = "2006"
}

@article{PhysRevD.109.024028,
  title = {Squeezing of light from Planck-scale physics},
  author = {Artigas, Danilo and Martineau, Killian and Mielczarek, Jakub},
  journal = {Phys. Rev. D},
  volume = {109},
  issue = {2},
  pages = {024028},
  numpages = {12},
  year = {2024},
  month = {Jan},
  publisher = {American Physical Society},
  doi = {10.1103/PhysRevD.109.024028},
  url = {https://link.aps.org/doi/10.1103/PhysRevD.109.024028}
}

@article{PhysRevD.104.126010,
  title = {Generalized uncertainty principle or curved momentum space?},
  author = {Wagner, Fabian},
  journal = {Phys. Rev. D},
  volume = {104},
  issue = {12},
  pages = {126010},
  numpages = {12},
  year = {2021},
  month = {Dec},
  publisher = {American Physical Society},
  doi = {10.1103/PhysRevD.104.126010},
  url = {https://link.aps.org/doi/10.1103/PhysRevD.104.126010}
}

@article{Nandi:2025mco,
    author = "Nandi, Partha and Roy, Mainak and Horoto, Langa and Scholtz, Frederik G. and Chakraborty, Biswajit",
    title = "{Quantum-gravitational backreaction in the BTZ background from curved momentum space}",
    eprint = "2509.05713",
    archivePrefix = "arXiv",
    primaryClass = "gr-qc",
    doi = "10.1088/1361-6382/ae62ef",
    journal = "Class. Quant. Grav.",
    volume = "43",
    number = "10",
    pages = "105015",
    year = "2026"
}

@article{PhysRevA.69.062321,
  title = {Minimal universal two-qubit controlled-NOT-based circuits},
  author = {Shende, Vivek V. and Markov, Igor L. and Bullock, Stephen S.},
  journal = {Phys. Rev. A},
  volume = {69},
  issue = {6},
  pages = {062321},
  numpages = {8},
  year = {2004},
  month = {Jun},
  publisher = {American Physical Society},
  doi = {10.1103/PhysRevA.69.062321},
  url = {https://link.aps.org/doi/10.1103/PhysRevA.69.062321}
}

@article{Maggiore1993,
  author  = {Maggiore, M.},
  title   = {A Generalized Uncertainty Principle in Quantum Gravity},
  journal = {Physics Letters B},
  volume  = {304},
  pages   = {65--69},
  year    = {1993},
  doi     = {10.1016/0370-2693(93)91401-8},
  eprint  = {hep-th/9301067},
  archivePrefix = {arXiv}
}

@article{PhysRevD.99.026012,
  title = {Modified commutation relationships from the Berry-Keating program},
  author = {Bishop, Michael and Aiken, Erick and Singleton, Douglas},
  journal = {Phys. Rev. D},
  volume = {99},
  issue = {2},
  pages = {026012},
  numpages = {7},
  year = {2019},
  month = {Jan},
  publisher = {American Physical Society},
  doi = {10.1103/PhysRevD.99.026012},
  url = {https://link.aps.org/doi/10.1103/PhysRevD.99.026012}
}

@article{Kempf1995,
author = {Kempf, Achim},
title = {On quantum field theory with nonzero minimal uncertainties in positions and momenta},
journal = {J. Math. Phys.},
volume = {38},
pages = {1347--1372},
year = {1997},
doi = {10.1063/1.531315}
}

@article{KempfManganoMann1995,
author = {Kempf, Achim and Mangano, Gianpiero and Mann, Robert B.},
title = {Hilbert space representation of the minimal length uncertainty relation},
journal = {Phys. Rev. D},
volume = {52},
pages = {1108--1118},
year = {1995},
doi = {10.1103/PhysRevD.52.1108}
}

@article{Hossenfelder2013,
author = {Hossenfelder, Sabine},
title = {Minimal Length Scale Scenarios for Quantum Gravity},
journal = {Living Rev. Relativity},
volume = {16},
pages = {2},
year = {2013},
doi = {10.12942/lrr-2013-2}
}

@article{JordanLeePreskill2012,
author = {Jordan, Stephen P. and Lee, Keith S. M. and Preskill, John},
title = {Quantum Algorithms for Quantum Field Theories},
journal = {Science},
volume = {336},
pages = {1130--1133},
year = {2012},
doi = {10.1126/science.1217069}
}

@article{JordanLeePreskill2014,
author = {Jordan, Stephen P. and Lee, Keith S. M. and Preskill, John},
title = {Quantum Computation of Scattering in Scalar Quantum Field Theories},
journal = {Quantum Inf. Comput.},
volume = {14},
pages = {1014--1080},
year = {2014}
}

@article{Klco_2019,
   title={Digitization of scalar fields for quantum computing},
   volume={99},
   ISSN={2469-9934},
   url={http://dx.doi.org/10.1103/PhysRevA.99.052335},
   DOI={10.1103/physreva.99.052335},
   number={5},
   journal={Physical Review A},
   publisher={American Physical Society (APS)},
   author={Klco, Natalie and Savage, Martin J.},
   year={2019},
   month=May }

@article{Feynman1982,
author = {Feynman, Richard P.},
title = {Simulating Physics with Computers},
journal = {Int. J. Theor. Phys.},
volume = {21},
pages = {467--488},
year = {1982},
doi = {10.1007/BF02650179}
}

@article{Lloyd1996,
author = {Lloyd, Seth},
title = {Universal Quantum Simulators},
journal = {Science},
volume = {273},
pages = {1073--1078},
year = {1996},
doi = {10.1126/science.273.5278.1073}
}

@book{Beauchamp1984,
author = {Beauchamp, Kenneth G.},
title = {Applications of Walsh and Related Functions},
publisher = {Academic Press},
year = {1984}
}

@article{Welch2014,
author = {Welch, Jonathan and Greenbaum, Daniel and Mostame, Sarah and Aspuru-Guzik, Alán},
title = {Efficient Quantum Circuits for Diagonal Unitaries Without Ancillas},
journal = {New J. Phys.},
volume = {16},
pages = {033040},
year = {2014},
doi = {10.1088/1367-2630/16/3/033040}
}

@article{ZylbermanDebbasch2024,
  author  = {Zylberman, Julien and Debbasch, Fabrice},
  title   = {Efficient quantum state preparation with {Walsh} series},
  journal = {Phys. Rev. A},
  volume  = {109},
  number  = {4},
  pages   = {042401},
  year    = {2024},
  doi     = {10.1103/PhysRevA.109.042401},
  eprint  = {2307.08384},
  archivePrefix = {arXiv},
  primaryClass  = {quant-ph}
}

@article{Bosso2023minimal,
  title   = {Minimal length: A cut-off in disguise?},
  author  = {Bosso, Pasquale and Petruzziello, Luciano and Wagner, Fabian},
  journal = {Phys. Rev. D},
  volume  = {107},
  issue   = {12},
  pages   = {126009},
  numpages = {10},
  year    = {2023},
  month   = {Jun},
  publisher = {American Physical Society},
  doi     = {10.1103/PhysRevD.107.126009},
  url     = {https://doi.org/10.1103/PhysRevD.107.126009}
}

@article{fine1949,
 ISSN = {00029947, 10886850},
 URL = {http://www.jstor.org/stable/1990619},
 author = {N. J. Fine},
 journal = {Transactions of the American Mathematical Society},
 number = {3},
 pages = {372--414},
 publisher = {American Mathematical Society},
 title = {On the Walsh Functions},
 urldate = {2026-07-13},
 volume = {65},
 year = {1949}
}

@ARTICLE{UpperBounds,
  author={Chung-Kwong Yuen},
  journal={IEEE Transactions on Computers}, 
  title={Upper Bounds on Walsh Transforms}, 
  year={1972},
  volume={C-21},
  number={12},
  pages={1273-1280},
  doi={10.1109/T-C.1972.223498}}

@misc{odonnellanalysisbooleanfunctions,
      title={Analysis of Boolean Functions}, 
      author={Ryan O'Donnell},
      year={2021},
      eprint={2105.10386},
      archivePrefix={arXiv},
      primaryClass={cs.DM},
      url={https://arxiv.org/abs/2105.10386}, 
}

@misc{bullock2003smallercircuitsarbitrarynqubit,
      title={Smaller Circuits for Arbitrary n-qubit Diagonal Computations}, 
      author={Stephen S. Bullock and Igor L. Markov},
      year={2003},
      eprint={quant-ph/0303039},
      archivePrefix={arXiv},
      primaryClass={quant-ph},
      url={https://arxiv.org/abs/quant-ph/0303039}, 
}

@misc{sinha2025lecturesquantumfieldtheory,
      title={Lectures on Quantum Field Theory on a Quantum Computer}, 
      author={Aninda Sinha and Ujjwal Basumatary},
      year={2025},
      eprint={2512.02706},
      archivePrefix={arXiv},
      primaryClass={quant-ph},
      url={https://arxiv.org/abs/2512.02706}, 
}

@article{Walsh1923,
  author  = {Walsh, Joseph L.},
  title   = {A Closed Set of Normal Orthogonal Functions},
  journal = {American Journal of Mathematics},
  volume  = {45},
  number  = {1},
  pages   = {5--24},
  year    = {1923},
  doi     = {10.2307/2387224}
}

@misc{kane2022efficientquantumimplementation21,
      title={Efficient quantum implementation of 2+1 U(1) lattice gauge theories with Gauss law constraints}, 
      author={Christopher Kane and Dorota M. Grabowska and Benjamin Nachman and Christian W. Bauer},
      year={2022},
      eprint={2211.10497},
      archivePrefix={arXiv},
      primaryClass={quant-ph},
      url={https://arxiv.org/abs/2211.10497}, 
}

@article{Hadfield:2018yak,
    author = "Hadfield, Stuart",
    title = "{On the Representation of Boolean and Real Functions as Hamiltonians for Quantum Computing}",
    eprint = "1804.09130",
    archivePrefix = "arXiv",
    primaryClass = "quant-ph",
    doi = "10.1145/3478519",
    journal = "ACM Trans. Quant. Comput.",
    volume = "2",
    number = "4",
    pages = "18",
    year = "2021"
}

@article{Bosso:2023aht,
    author = "Bosso, Pasquale and Luciano, Giuseppe Gaetano and Petruzziello, Luciano and Wagner, Fabian",
    title = "{30 years in: Quo vadis generalized uncertainty principle?}",
    eprint = "2305.16193",
    archivePrefix = "arXiv",
    primaryClass = "gr-qc",
    doi = "10.1088/1361-6382/acf021",
    journal = "Class. Quant. Grav.",
    volume = "40",
    number = "19",
    pages = "195014",
    year = "2023"
}

@article{Banerjee:2010sd,
    author = "Banerjee, Rabin and Ghosh, Sumit",
    title = "{Generalised Uncertainty Principle, Remnant Mass and Singularity Problem in Black Hole Thermodynamics}",
    eprint = "1002.2302",
    archivePrefix = "arXiv",
    primaryClass = "gr-qc",
    doi = "10.1016/j.physletb.2010.04.008",
    journal = "Phys. Lett. B",
    volume = "688",
    pages = "224--229",
    year = "2010"
}

@article{Bosso:2020aqm,
    author = "Bosso, Pasquale",
    title = "{On the quasi-position representation in theories with a minimal length}",
    eprint = "2005.12258",
    archivePrefix = "arXiv",
    primaryClass = "gr-qc",
    doi = "10.1088/1361-6382/abe758",
    journal = "Class. Quant. Grav.",
    volume = "38",
    number = "7",
    pages = "075021",
    year = "2021"
}

@article{PhysRevD.52.1108,
  title = {Hilbert space representation of the minimal length uncertainty relation},
  author = {Kempf, Achim and Mangano, Gianpiero and Mann, Robert B.},
  journal = {Phys. Rev. D},
  volume = {52},
  issue = {2},
  pages = {1108--1118},
  numpages = {0},
  year = {1995},
  month = {Jul},
  publisher = {American Physical Society},
  doi = {10.1103/PhysRevD.52.1108},
  url = {https://link.aps.org/doi/10.1103/PhysRevD.52.1108}
}
\end{document}